\documentclass[12pt,english,transactions]{IEEEtran}
\usepackage[LGR,T1]{fontenc}
\usepackage[latin9]{inputenc}
\usepackage{geometry}
\usepackage{float}
\usepackage{amsmath}
\usepackage{amsthm}
\usepackage{amssymb}
\usepackage{graphicx}
\usepackage{setspace}
\PassOptionsToPackage{normalem}{ulem}
\usepackage{ulem}
\makeatletter

\DeclareRobustCommand{\greektext}{%
  \fontencoding{LGR}\selectfont\def\encodingdefault{LGR}}
\DeclareRobustCommand{\textgreek}[1]{\leavevmode{\greektext #1}}
\DeclareFontEncoding{LGR}{}{}
\DeclareTextSymbol{\~}{LGR}{126}
\providecommand{\tabularnewline}{\\}

\theoremstyle{plain}
\newtheorem{thm}{\protect\theoremname}
\theoremstyle{definition}
\newtheorem{example}[thm]{\protect\examplename}
\theoremstyle{definition}
\newtheorem{defn}[thm]{\protect\definitionname}

 \date{ }

\renewcommand\thesection{\roman{section}}
\onecolumn

\makeatother

\usepackage{babel}
\providecommand{\definitionname}{Definition}
\providecommand{\examplename}{Example}
\providecommand{\theoremname}{Theorem}

\begin{document}

\title{Interference Management in Heterogeneous Networks with Blind Transmitters}

\author{Vaia Kalokidou\footnote{Vaia Kalokidou and Robert Piechocki are with the Communication Systems and Networks Research Group at Department of Electrical and Electronic Engineering (MVB), School of Engineering, University of Bristol, UK.},
Oliver Johnson\footnote{Oliver Johnson is with the Department of Mathematics, University of Bristol, UK.},
Robert Piechocki}
\maketitle
\begin{abstract}
\begin{onehalfspace}
Future multi-tier communication networks will require enhanced network
capacity and reduced overhead. In the absence of Channel State Information
(CSI) at the transmitters, Blind Interference Alignment (BIA) and
Topological Interference Management (TIM) can achieve optimal Degrees
of Freedom (DoF), minimizing network's overhead. In addition, Non-Orthogonal
Multiple Access (NOMA) can increase the sum rate of the network, compared
to orthogonal radio access techniques currently adopted by 4G networks.
Our contribution is two interference management schemes, BIA and a
hybrid TIM-NOMA scheme, employed in heterogeneous networks by applying
user-pairing and Kornecker Product representation. BIA manages inter-
and intra-cell interference by antenna selection and appropriate message
scheduling. The hybrid scheme manages intra-cell interference based
on NOMA and inter-cell interference based on TIM. We show that both
schemes achieve at least double the rate of TDMA. The hybrid scheme
always outperforms TDMA and BIA in terms of Degrees of Freedom (DoF).
Comparing the two proposed schemes, BIA achieves more DoF than TDMA
under certain restrictions, and provides better Bit-Error-Rate (BER)
and sum rate perfomance to macrocell users, whereas the hybrid scheme
improves the performance of femtocell users. 
\end{onehalfspace}

\renewcommand\thesection{\roman{section}}
\end{abstract}

\section{Introduction}

Over the past few years, cellular and wireless networks have been
challenged by the increasing number of mobile Internet services and
the constant growth of mobile data traffic. Future radio access networks
should provide reduced latency, improved energy efficiency, and high
user data rates in dense and high mobility network environments. The
architecture of future communication networks will be heterogeneous
in nature, i.e. macrocell with many small cells. The great challenge
will be the employment of novel interference management strategies
that will manage interference, without increasing the system's overhead,
and provide high data rates and reliable transmissions.

Interference Alignment (IA) was introduced by Maddah-Ali et al. in
\cite{IAXChannel-1}, and Jafar and Shamai in \cite{IAXChannel-2}
for the MIMO X channels, and by Cadambe and Jafar in \cite{IAIntChannel-1}
for the $K$-user interference channel, where $K/2$ Degrees of Freedom
(DoF) can be achieved. IA aligns the interfering signals present at
each receiver into a low dimensional subspace, by linearly encoding
signals in multiple dimensions, resulting in the desired signal being
in a dimension unoccupied by interference links. Initially, IA required
global Channel State Information (CSI) and was computationally complex. 

Further work on IA led to the scheme of Blind IA, presented by Wang,
Gou and Jafar in \cite{BIA 1} and Jafar in \cite{BIA 2}, for certain
network scenarios, which can achieve full DoF in the absence of CSI
at the transmitters (CSIT), thus reducing the system overhead. Furthermore,
Blind IA was introduced, by Jafar in \cite{CellularBIA}, for cellular
and heterogeneous networks, by ``seeing'' frequency reuse (i.e.
orthogonal allocation of signaling dimensions) as a simple form of
interference alignment. Blind IA in heterogeneous networks was generalized
in \cite{Paper1} for the case of $K$ users in the macrocell and
$K$ femtocells with one user each, introducing Kronecker (Tensor)
Product representation and a variation of model parameters to optimize
the sum rate performance. A special case of Blind IA, known as Topological
Interference Management (TIM), was introduced by Jafar in \cite{TIM}.
TIM takes into consideration the position of every user in the cell(s),
and based on their channel strength, weak interference links are ignored,
resulting in $1/2$ DoF achieved for every user in the SISO Broadcast
Channel (BC). In \cite{TIM-MIMO}, Sun and Jafar research the scheme
of TIM for the case of multiple receive and transmit antennas, concluding
that only the former can provide more DoF in the network. 

Unlike Orthogonal Frequency Division Multiple Access (OFDMA) and Single-Carrier
Frequency Division Multiple Access (SC-FDMA) currently employed in
4G mobile networks, the scheme of Non-Orthogonal Multiple Access (NOMA),
proposed in \cite{NOMA1} by Saito et al., is based on a non-orthogonal
approach to future radio access. According to NOMA, multiple users
are superimposed in the power domain at the transmitters, and Successive
Interference Cancellation (SIC) is performed at the receivers, improving
capacity and throughput performance. 

Power allocation and Quality-of-Service (QoS) for edge-cell users,
has been a major issue to tackle in systems employing NOMA. It has
been shown, in \cite{NOMA8}, that for the BC, if NOMA is employed
with the aid of Coordinated Multiple Point (CoMP) and\emph{ }Alamouti
Code, satisfactory rates for edge-cell users can be achieved without
degrading the performance of users' closer to the base station. Moreover,
with an adaptive power and frequency resource allocation algorithm,
as proposed in \cite{NOMA9}, targeting inter-cell interference, in
order to boost the total throughput, reliable transmissions to edge-cell
users can be obtained. Furthermore, in \cite{NOMAnew}, authors study
two different power allocation schemes, a fixed one and a cognitive
radio inspired one, in a MIMO-NOMA model by using signal alignment
and stochastic geometry.

Recently, research on NOMA has been focusing on user pairing to reduce
complexity and improve efficiency. Cooperative NOMA schemes, where
users with higher channel gains have prior information about other
users' messages, have been developed \cite{NOMA13}. User pairing
has been introduced in two NOMA schemes, discussed in \cite{NOMA21},
with one scheme employing fixed power allocation (F-NOMA) and another
one inspired by cognitive radio (CR-NOMA), with users grouped differently
in each one of the two NOMA schemes. In addition, user pairing has
been studied in conjunction with the problem of power allocation,
in \cite{NOMA22}, based on a new design of precoding and detection
matrices. User pairing and the performance of NOMA have been also
studied from an information theory perspective, as discussed in \cite{NOMA24},
researching the relationship between the rate region achieved by NOMA
and the capacity region of the BC, observing that different power
allocation to users corresponds to different points on the rate region
graph, and showing that NOMA can outperform TDMA not only in terms
of the sum rate, but for every users's rate as well. 

From the schemes of TIM and NOMA, a hybrid TIM-NOMA scheme emerged,
introduced in \cite{Paper2} for the SISO BC and in \cite{Paper3}
for the MIMO BC. The hybrid TIM-NOMA scheme divides users into groups,
and manages ``inter-group'' interference based on the principles
of TIM, and ``intra-group'' interference based on NOMA. The hybrid
scheme can achieve double the sum rate of TDMA for high SNR values.

\begin{figure}
\begin{centering}
\begin{minipage}[t]{0.45\textwidth}%
\begin{center}
\includegraphics[width=6cm,height=5cm]{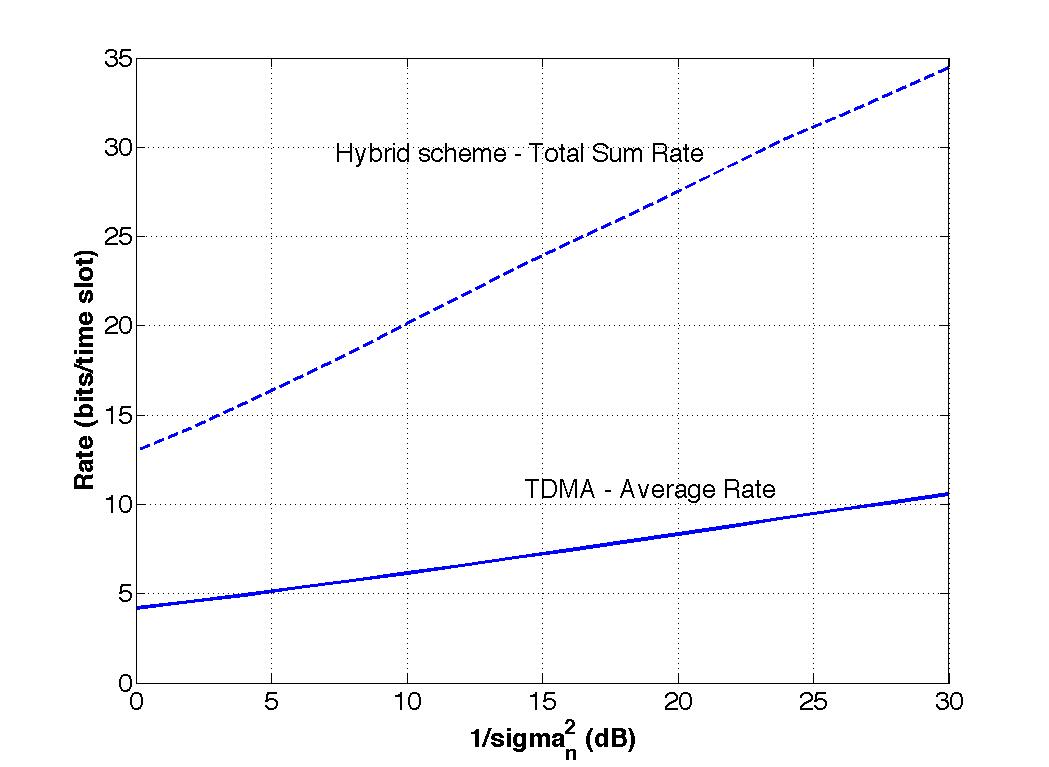}
\par\end{center}%
\end{minipage}\thinspace %
\begin{minipage}[t]{0.45\textwidth}%
\begin{center}
\includegraphics[width=6cm,height=5cm]{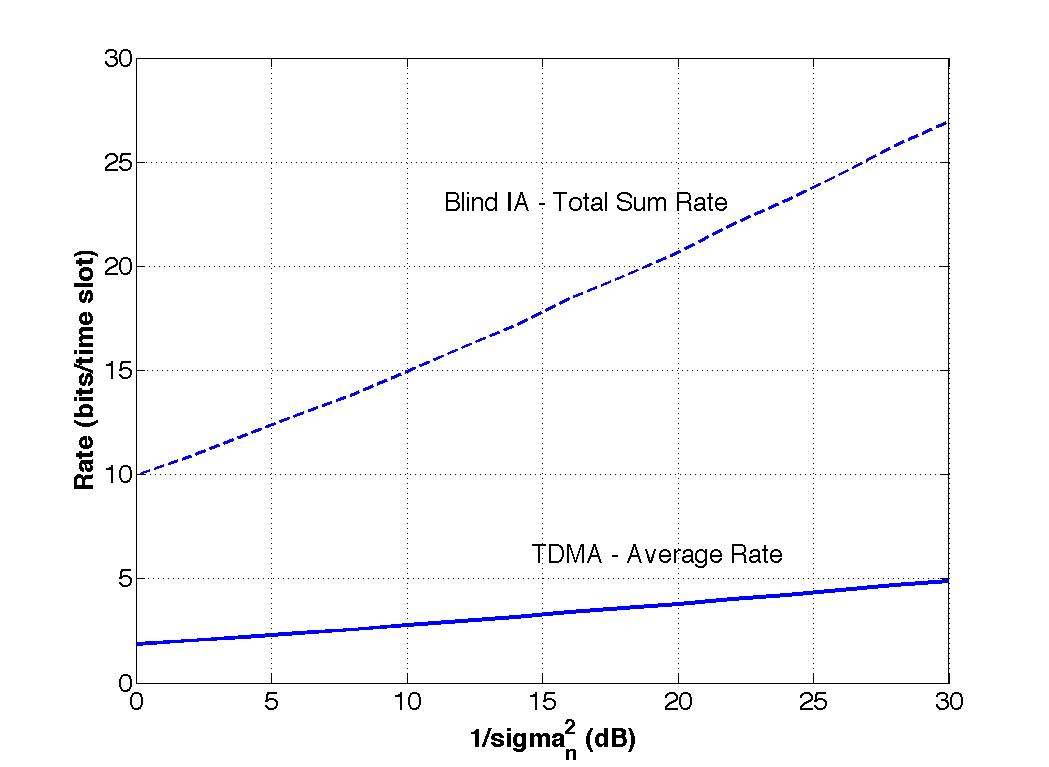}
\par\end{center}%
\end{minipage}
\par\end{centering}

\caption{Sum Rate Performance: (left) Hybrid vs. TDMA, (right) Blind IA vs.
TDMA. Both schemes achieve higher sum rates than TDMA \label{fig:Journal1-1-1-2-1-1-1-1-1-1} }
\end{figure}

In this paper, based on \cite{Paper1,Paper2,Paper3}, we introduce
two interference management schemes employed in heterogeneous networks.
For both schemes, we consider a $K$-user macrocell and $KL$ femtocells
with one user each, taking into consideration the position of every
user in the cell. The first scheme is a hybrid TIM-NOMA scheme based
on \cite{Paper2,Paper3}. The novelty of this scheme is the fact that
it changes the way user-grouping is performed compared to \cite{Paper2,Paper3}.
Users in the macrocell belong to one group and then there exist $L$
groups of femtocells. Inter-cell interference is managed based on
TIM and intra-cell interference based on NOMA. The second scheme is
Blind IA in heterogeneous networks, which constitutes further work
on \cite{Paper1}. Our contribution is the additional consideration
of interference caused to femtocells by transmissions in the macrocell,
and the existence of more than one femtocells around a macrocell user.
The algorithms of both schemes are described by using Kronecker (Tensor)
product representation. Based on our results, the hybrid scheme can
achieve more total DoF compared to TDMA, whereas Blind IA outperforms
TDMA in terms of DoF in most cases. We show that both schemes achieve
higher sum rates than TDMA, as depicted in Figure \ref{fig:Journal1-1-1-2-1-1-1-1-1-1}.
Finally, comparing the two schemes, Blind IA provides better sum-rate
and BER performance to macrocell users, whereas the hybrid scheme
results in better performance for the users in the femtocells, and
based on its power allocation scheme provides QoS to edge cell users.

The rest of the paper is organized as follows. Section 2 describes
the general network architecture, and the example-model which is used
to describe the two schemes. Sections 3 and 4 present the model description
and achievable sum rate of the hybrid and Blind IA schemes respectively.
Section 5 describes the special case of Blind IA when $L=1$, i.e.
only one femtocell interferes with every user in the macrocell. Section
6 discusses our results and the performance of the two schemes in
terms of DoF, BER and sum rate. Finally, Section 7 summarizes the
main findings of our work and discusses further developments.

\section{System Model }

\begin{figure}
\begin{centering}
\includegraphics[width=0.45\columnwidth]{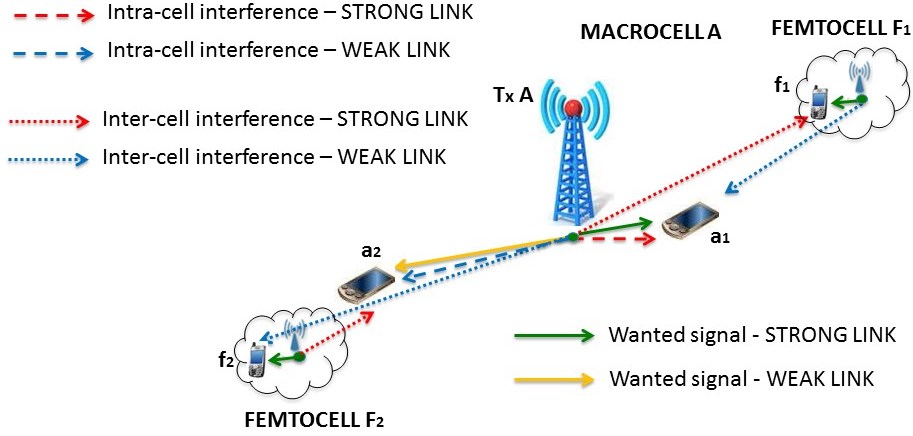}
\par\end{centering}

\caption{Heterogeneous network: $K=2$ users in macrocell and $KL=2$ femtocell
($L=1$) with 1 user each ($N=2$ transmit and receive antennas).\label{fig:Heterogeneous-network:-}}
\end{figure}

Consider the Broadcast Channel of a heterogeneous network, as shown
in Figure \ref{fig:Heterogeneous-network:-}, with 1 macrocell and
$KL$ femtocells. At the \emph{$N\times N$ }MIMO BC of the\emph{
macrocell}, there is one transmitter \emph{$T_{xA}$ }with \emph{$N$}
antennas, and \emph{$K$} users equipped with \emph{$N$} antennas
each. Transmitter \emph{$T_{xA}$ }has $N$ messages to send to every
user, and when it transmits to user \emph{$a_{k}$}, where $k\in\left\{ 1,2,...,K\right\} $,
it causes interference to the other $K-1$ users in the macrocell
and all the femtocell users $f_{kl}$. $L$ femtocells are considered
to interfere with every macrocell user. At the \emph{$N\times N$
}MIMO BC of each \emph{femtocell}, there is one transmitter \emph{$T_{xkl}$}
with $N$ antennas, and one user $f_{kl}$ equipped with $N$ antennas.
When transmitter \emph{$T_{xkl}$} transmits to user $f_{kl}$, it
causes interference to the macrocell user\emph{ $a_{k}$} and to all
or some (depending on the scheme we use) of the remaining $L-1$ neighbouring
femtocell users $f_{kl}$. We consider that all channels remain constant
over $T$ time slots (i.e. supersymbol) and we take into consideration
the position of users in the cells, as summarized in Table \ref{tab:Ch7-Distance-metrics-in}. 

In the hybrid TIM-NOMA scheme, users are divided into $G=T=L+1$ groups.
In the macrocell, all users belong to the same group $G_{0}$. In
addition, there are $L$ different groups of femtocell users, with
$G_{l}=\left\{ f_{1l},f_{2l},...,f_{Kl}\right\} $ for $l\in\left\{ 1,2,...,L\right\} $.
Moreover, $N$ messages are transmitted to every user in the femtocells.

In the Blind IA scheme, the number of neighbouring femtocells cannot
be greater than 3, i.e. $L\leqslant3$, with $L\leqslant N$. There
is no grouping in the macrocell. Every macrocell user receives interference
from $L$ femtocells. Thus, for all $k$, with $k\in\left\{ 1,2,...,K\right\} $,
femtocell users are divided into two groups: Group $G_{1}$ consists
of $(L-1)$ femtocell users $f_{kl}$, i.e. $G_{1}=\left\{ f_{1l},f_{2l},...,f_{Kl}\right\} $,
where $l\in\left\{ 1,...,L\right\} $ and $l\neq2$. Group $G_{2}$
consists of femtocell users $f_{kl}$, i.e. $G_{2}=\left\{ f_{1l},f_{2l},...,f_{Kl}\right\} $,
where $l=2$. Femtocells in $G_{1}$ do not interfere with each other.
Femtocells in $G_{2}$ interfere with all femtocells in $G_{1}$.
In addition, $\mathcal{M}_{1}=\left(K-1\right)\left(N-\left(L-1\right)\right)$
messages are sent to users in $G_{1}$, and $\mathcal{M}_{2}=1$ messages
are sent to users in $G_{2}$.

In this paper, we consider the following example model: In the macrocell,
there are $K=2$ users, $N=2$ messages intended for every user, and
\emph{$N=2$} transmit and receive antennas. Additionally there are
$KL=2$ femtocells (note that $L=1$), with \emph{$N=2$} transmit
and receive antennas\emph{ }each. For the hybrid scheme, we consider
$N=2$ messages sent to every femtocell user, $T=2$ time slots and
$G=2$ groups ($\left\{ G_{0},G_{1}\right\} $), and for the Blind
IA scheme, $\mathcal{M}_{1}=2$ messages sent to users in $G_{1}$,
$T=3$ time slots and $G=1$ groups ($\left\{ G_{1}\right\} $).

\begin{table}[H]
\begin{centering}
\begin{tabular}{|c|c|}
\hline 
\textbf{\footnotesize{}Description} & \textbf{\footnotesize{}Value}\tabularnewline
\hline 
\hline 
{\footnotesize{}Macrocell Radius} & {\footnotesize{}$5\,km$}\tabularnewline
\hline 
{\footnotesize{}Femtocell Radius} & {\footnotesize{}$0.2\,km$}\tabularnewline
\hline 
{\footnotesize{}Reuse Distance Macrocell-Femtocell $F_{1}$} & {\footnotesize{}$5\,km$}\tabularnewline
\hline 
{\footnotesize{}Reuse Distance Macrocell-Femtocell $F_{2}$} & {\footnotesize{}$5.2\,km$}\tabularnewline
\hline 
{\footnotesize{}Distance of $a_{1}$ from $T_{x_{A}}$} & {\footnotesize{}$0.5\,km$}\tabularnewline
\hline 
{\footnotesize{}Distance of $a_{2}$ from $T_{x_{A}}$} & {\footnotesize{}$4.5\,km$}\tabularnewline
\hline 
{\footnotesize{}Distance of $f_{1}$ from $T_{x_{F1}}$} & {\footnotesize{}$0.2\,km$}\tabularnewline
\hline 
{\footnotesize{}Distance of $f_{2}$ from $T_{x_{F2}}$} & {\footnotesize{}$0.2\,km$}\tabularnewline
\hline 
\end{tabular}
\par\end{centering}

\caption{Distance metrics for the example model heterogeneous network.\label{tab:Ch7-Distance-metrics-in}}
\end{table}

\section{Hybrid TIM-NOMA scheme}

\uline{In the macrocell,} the $TN\times1$ signal at receiver $a_{k}$,
considering slow fading (i.e. channels are fixed through transmission
time), is given by:

\begin{equation}
\mathbf{y}_{a_{k}}=\mathbf{H}_{a_{k}}\mathbf{x}_{A}+\sum_{l=1}^{L}\mathbf{H}_{f_{kl}a_{k}}\mathbf{x}_{kl}+\mathbf{z}_{a_{k}},\label{eq:MacrocellSignal}
\end{equation}

where $\mathbf{H}_{a_{k}}\in\mathbb{C}^{(TN\times TN)}$ is the channel
transfer matrix from $T_{x_{A}}$ to receiver $a_{k}$ and is given
by $\mathbf{H}_{a_{k}}=\sqrt{\gamma_{a_{k}}}(\mathbf{I}_{T}\otimes\mathbf{h}_{a_{k}})$
(here and throughout $\mathbf{H}_{k}=\sqrt{\gamma_{k}}(\mathbf{I}_{T}\otimes\mathbf{h}_{k})$
with $\mathbf{h}_{k}$ denoting the channel coefficients from $T_{x_{K}}$
to $k$ for one time slot, $\gamma_{k}=1/d_{k}^{n}$ the path loss
of user $k$ with $n$ denoting the path loss exponent considered
for an urban environment ($n=3$), and $\otimes$ the Kronecker (Tensor)
product). $\mathbf{H}_{f_{kl}a_{k}}\in\mathbb{C}^{(TN\times TN)}$
is the interference channel transfer matrix from $T_{x_{kl}}$ to
receiver $a_{k}$ (here and throughout $\mathbf{H}_{xk}=\sqrt{\gamma_{xk}}(\mathbf{I}_{T}\otimes\mathbf{h}_{xk})$
with $\mathbf{h}_{xk}$ denoting the interference channel coefficients
from $T_{x_{K}}$ to $k$ for one time slot). Due to the users' different
locations, channel coefficients are statistically independent, and
follow an i.i.d. Gaussian distribution $\mathcal{CN}(0,1)$. Finally,
$\mathbf{z}_{a_{k}}\sim\mathcal{CN}(0,\text{\textgreek{sv}}_{n}^{2}\mathbf{I}_{TN})$
denotes the independent Additive White Gaussian Noise (AWGN) at the
input of receiver $a_{k}$.

Taking into consideration the position of each user $a_{k}$ in the
macrocell, users are ordered increasingly, in increasing order of
path loss $\gamma_{a_{k}}$.
\begin{example}
For the example model, assuming both macrocell users have the same
received noise power $\text{\textgreek{sv}}_{n}^{2}$, it follows:
\begin{equation}
\gamma_{1}>\gamma_{2},\label{eq:Ch7-MacroPL}
\end{equation}
with user 1 being very close to the base station and user $2$ at
the edge of the cell. Weaker channels, of users' being far from the
base station, need to be boosted, such that for the transmit power
$P_{k}$ of every user it holds that $P_{2}>P_{1}$. For every user
$a_{k}$ in the macrocell, we choose to take their transmitted power,
as initially suggested in \cite{Paper2}, given by:
\begin{equation}
P_{a_{k}}=\frac{a^{2}}{N}\frac{d_{a_{k}}^{2}}{\sum_{j=1}^{K}d_{a_{j}}^{2}},
\end{equation}

\end{example}
where $a\in\mathbb{R}$ is a constant determined by power considerations
(see (\ref{eq:Ch7-femtosa})). The total transmit power in the macrocell
is given by the power constraint:
\begin{equation}
P_{macrocell}=\sum_{j=1}^{K}d_{a_{j}}^{2}P_{a_{j}}\mathrm{norm}(\mathbf{u}_{a_{j}})=a^{2}.\label{eq:Ch7-femtosa}
\end{equation}

Then, the $T\times1$ transmitted vector $\mathbf{x}_{A}$ is given
by:

\begin{equation}
\mathbf{x}_{A}=\left(\mathbf{v}_{0}\otimes\mathbf{I}_{N}\right)\sum_{k=1}^{K}\sqrt{P_{a_{k}}}\mathbf{u}_{a_{k}},
\end{equation}

with $\mathbf{v}_{0}$ denoting the $T\times1$ precoding vector corresponding
to group $G_{0}$ that macrocell users belong to, and should be orthogonal
to all the remaining $T-1$ precoding vectors (corresponding to $T-1$
groups). 

\uline{In each femtocell,} the $TN\times1$ signal at receiver
$f_{kl}$ is given by:

\begin{equation}
\mathbf{y}_{f_{kl}}=\mathbf{H}_{f_{kl}}\mathbf{x}_{f_{kl}}+\mathbf{H}_{Af_{kl}}\mathbf{x}_{A}+\sum_{\overset{j=1}{j\neq l}}^{L}\mathbf{H}_{f_{kj}f_{kl}}\mathbf{x}_{kj}+\mathbf{z}_{f_{kl}},
\end{equation}

where $\mathbf{H}_{f_{kl}}\in\mathbb{C}^{(TN\times TN)}$ is the channel
transfer matrix from $T_{xkl}$ to receiver $f_{kl}$, $\mathbf{H}_{Af_{kl}}\in\mathbb{C}^{(TN\times TN)}$
is the channel transfer matrix from $T_{x_{A}}$ to receiver $f_{kl}$,
and $\mathbf{H}_{f_{kj}f_{kl}}\in\mathbb{C}^{(TN\times TN)}$ is the
channel transfer matrix from $T_{xkj}$ to receiver $f_{kl}$. Finally,
$\mathbf{z}_{f_{kl}}\sim\mathcal{CN}(0,\text{\textgreek{sv}}_{n}^{2}\mathbf{I}_{TN})$
denotes the independent AWGN at the input of receiver $f_{kl}$.

For user $f_{kl}$ in every femtocell, their transmitted power is
given by $P_{f_{kl}}=b^{2}/N$, where $b\in\mathbb{R}$ is a constant
determined by power considerations (see (\ref{eq:Ch7-femtosb})),
and the total transmit power in the femtocell is given by:

\begin{equation}
P_{f_{emtocell}}=b^{2}.\label{eq:Ch7-femtosb}
\end{equation}

Then, the $T\times1$ transmitted vector $\mathbf{x}_{f_{kl}}$ is
given by:

\begin{equation}
\mathbf{x}_{f_{kl}}=\left(\mathbf{v}_{l}\otimes\mathbf{I}_{N}\right)\sqrt{P_{f_{kl}}}\mathbf{u}_{f_{kl}},
\end{equation}

with $\mathbf{v}_{l}$ denoting the $T\times1$ precoding unit vector
corresponding to group $G_{l}$ that user $f_{kl}$ belongs to, and
should be orthonormal to all the remaining $T-1$ precoding vectors. 
\begin{example}
For the example model, we choose the precoding vectors $\mathbf{v}_{0}$
and $\mathbf{v}_{1}$, for groups $G_{0}$ and $G_{1}$ respectively,
as $\mathbf{v}_{0}=\begin{bmatrix}1/2 & \sqrt{3}/2\end{bmatrix}^{T}$,$\mathbf{v}_{1}=\begin{bmatrix}-\sqrt{3}/2 & 1/2\end{bmatrix}^{T}$,
where $G_{0}=\left\{ a_{1},a_{2}\right\} $ and $G_{1}=\left\{ f_{\text{11}},f_{21}\right\} $.
\end{example}

\subsection{Inter-cell Interference Management}

In the network, there will be one $T\times1$ unit precoding vector
$\mathbf{v}_{0}$ for the macrocell and ($T-1$) $T\times1$ unit
precoding vectors $\mathbf{v}_{l}$, where $l\in\left\{ 1,2,...,L\right\} $,
for the femtocells, with all precoding vectors being orthogonal to
each other. 
\begin{thm}
In the macrocell, multiplying the received signal $\mathbf{y}_{a_{k}}$
with $\mathbf{v}_{0}{}^{T}\otimes\mathbf{I}_{N}$, the resulting signal
at every receiver $a_{k}$, is given by:\emph{
\begin{equation}
\widetilde{\mathbf{y}}_{a_{k}}=\left(\sum_{j=1}^{K}\sqrt{P_{a_{j}}}\sqrt{\gamma_{a_{k}}}\mathbf{h}_{a_{k}}\mathbf{u}_{a_{j}}\right)+\widetilde{\mathbf{z}_{a_{k}}},
\end{equation}
}where $\widetilde{\mathbf{z}_{a_{k}}}=\left(\mathbf{v}_{0}{}^{T}\otimes\mathbf{I}_{N}\right)\mathbf{z}_{a_{k}}$
remains white noise with the same variance.\end{thm}
\begin{IEEEproof}
We show that $\left(\mathbf{v}_{0}{}^{T}\otimes\mathbf{I}_{N}\right)$
removes inter-cell interference at the $k$th receiver $a_{k}$: 
\begin{align}
\left(\mathbf{v}_{0}^{T}\otimes\mathbf{I}_{N}\right)\mathbf{y}_{a_{k}} & =\left(\mathbf{v}_{0}^{T}\otimes\mathbf{I}_{N}\right)\left(\sqrt{\gamma_{a_{k}}}(\mathbf{I}_{T}\otimes\mathbf{h}_{a_{k}})\left(\mathbf{v}_{0}\otimes\mathbf{I}_{N}\right)\sum_{k=1}^{K}\sqrt{P_{a_{k}}}\mathbf{u}_{a_{k}}\right)\nonumber \\
 & +\left(\mathbf{v}_{0}^{T}\otimes\mathbf{I}_{N}\right)\left(\sum_{l=1}^{L}\sqrt{\gamma_{f_{kl}a_{k}}}(\mathbf{I}_{T}\otimes\mathbf{h}_{f_{kl}a_{k}})\left(\mathbf{v}_{l}\otimes\mathbf{I}_{N}\right)\sqrt{P_{f_{kl}}}\mathbf{u}_{f_{kl}}\right)+\widetilde{\mathbf{z}_{a_{k}}}\nonumber \\
 & =\sum_{j=1}^{K}\sqrt{P_{a_{j}}}\sqrt{\gamma_{a_{k}}}\mathbf{h}_{a_{k}}\mathbf{u}_{a_{j}}+\sum_{l=1}^{L}\sqrt{\gamma_{f_{kl}a_{k}}}\left(\mathbf{v}_{0}^{T}\mathbf{v}_{l}\otimes\mathbf{h}_{f_{kl}a_{k}}\right)\sqrt{P_{f_{kl}}}\mathbf{u}_{f_{kl}}+\widetilde{\mathbf{z}_{a_{k}}},
\end{align}

where by definition, for $l=1,\ldots,L$, $\mathbf{v}_{0}{}^{T}\mathbf{v}_{l}=0$. \end{IEEEproof}
\begin{thm}
In the femtocell, multiplying the received signal $\mathbf{y}_{f_{kl}}$
with $\mathbf{v}_{l}{}^{T}\otimes\mathbf{I}_{N}$, the resulting signal
at every receiver $f_{kl}$, is given by:\emph{
\begin{equation}
\widetilde{\mathbf{y}}_{f_{kl}}=\sqrt{\gamma_{f_{kl}}}\sqrt{P_{f_{kl}}}\mathbf{h}_{f_{kl}}\mathbf{u}_{f_{kl}}+\widetilde{\mathbf{z}_{f_{kl}}},
\end{equation}
}where $\widetilde{\mathbf{z}_{f_{kl}}}=\left(\mathbf{v}_{l}{}^{T}\otimes\mathbf{I}_{N}\right)\mathbf{z}_{f_{kl}}$
remains white noise with the same variance.\end{thm}
\begin{IEEEproof}
We show that $\left(\mathbf{v}_{l}{}^{T}\otimes\mathbf{I}_{N}\right)$
removes inter-cell interference at the $kl$th receiver $f_{kl}$:
\begin{align}
\left(\mathbf{v}_{l}^{T}\otimes\mathbf{I}_{N}\right)\mathbf{y}_{f_{kl}} & =\sqrt{\gamma_{f_{kl}}}\sqrt{P_{f_{kl}}}\mathbf{h}_{f_{kl}}\mathbf{u}_{f_{kl}}+\sqrt{\gamma_{Af_{kl}}}\left(\mathbf{v}_{l}{}^{T}\mathbf{v}_{0}\otimes\mathbf{h}_{Af_{kl}}\right)\sum_{j=1}^{K}\sqrt{P_{a_{j}}}\mathbf{u}_{a_{j}}\nonumber \\
 & +\sum_{\overset{j=1}{j\neq l}}^{L}\sqrt{\gamma_{f_{kj}f_{kl}}}\left(\mathbf{v}_{l}^{T}\mathbf{v}_{j}\otimes\mathbf{h}_{f_{kj}f_{kl}}\right)\sqrt{P_{f_{kj}}}\mathbf{u}_{f_{kj}}+\widetilde{\mathbf{z}_{f_{kl}}},
\end{align}

where by definition, for $l=1,\ldots,L$, $\mathbf{v}_{0}{}^{T}\mathbf{v}_{l}=0$,
and for $j=1,\ldots,L$ and $j\neq l$, $\mathbf{v}_{l}{}^{T}\mathbf{v}_{j}=0$.\end{IEEEproof}
\begin{example}
For the example model, for groups $G_{0}$ and $G_{1}$, the post-processed
signals at receivers $a_{1}$ and $f_{1}$ are: \emph{
\[
\widetilde{\mathbf{y}_{a_{1}}}=\sum_{j=1}^{2}\sqrt{P_{a_{j}}}\sqrt{\gamma_{a_{1}}}\mathbf{h}_{a_{1}}\mathbf{u}_{a_{j}}+\widetilde{\mathbf{z}_{a_{1}}},
\]
\[
\widetilde{\mathbf{y}_{f_{1}}}=\sqrt{\gamma_{f_{1}}}\sqrt{P_{f_{1}}}\mathbf{h}_{f_{1}}\mathbf{u}_{f_{1}}+\widetilde{\mathbf{z}_{f_{1}}}.
\]
}
\end{example}

\subsection{Intra-cell Interference Management}

The concept of NOMA will be only applied to group $G_{0}$ since only
users in the macrocell experience intra-cell interference. Users are
ordered in increasing order of their path loss $\gamma_{a_{k}}$ and
SIC is performed at every receiver. Each user $a_{k}$ can correctly
decode the signals of users whose path loss is smaller than theirs
by considering their own signal as noise. In the case that $a_{k}$
receives interference from transmissions to users in the macrocell
that have a larger path loss than they do, then $a_{k}$ decodes their
own signal considering interference as noise. Maximum Likelihood (ML)
reception is performed every time a user decodes its own or another
user's signal.
\begin{example}
The decoding order for the macrocell users is given in (\ref{eq:Ch7-MacroPL}).
Receiver $a_{2}$ decodes their own signal, considering interference
from transmissions to user $a_{1}$ as noise. Receiver $a_{1}$ decodes
first signal $\mathbf{u}_{a_{2}}$ (finding $\widetilde{\mathbf{u}_{a_{2}}}$),
considering their own signal as noise, and subtracts the estimate
$\widetilde{\mathbf{u}_{a_{2}}}$ from their post-processed signal
$\widetilde{\mathbf{y}_{a_{1}}}$. Then, they decode their own signal
as $\widetilde{\widetilde{\mathbf{y}_{a_{1}}}}=\widetilde{\mathbf{y}_{a_{1}}}-\sqrt{\gamma_{a_{1}}}\sqrt{P_{a_{1}}}\mathbf{h}_{a_{1}}\widetilde{\mathbf{u}_{a_{2}}}$,
which if $\widetilde{\mathbf{u}_{a_{2}}}=\mathbf{u}_{a_{2}}$ reduces
to the interference-free channel.
\end{example}

\subsection{Achievable Sum Rate}

In the macrocell, the total rate for each user $a_{k}$ per time slot,
setting $S=\left({\displaystyle \sum_{\overset{j\in G_{0}}{j<k}}}\left|\mathbf{h}_{a_{k}}\right|^{2}P_{a_{j}}+\sigma_{n}^{2}\right)$,
is given by:

\begin{equation}
R_{a_{k}}=\frac{1}{T}\mathbb{E}\left[\log\det\left(\mathbf{I}_{N}+\frac{P_{macrocell}}{NS}\frac{d_{a_{k}}^{2}}{{\displaystyle \sum_{j=1}^{K}}d_{a_{j}}^{2}}\gamma_{a_{k}}\mathbf{h}_{a_{k}}\mathbf{h}_{a_{k}}^{T}\right)\right],
\end{equation}

where $a_{k}\in G_{0}$.

In every femtocell, the total rate for each user $f_{kl}$, per time
slot is given by:

\begin{equation}
R_{f_{kl}}=\frac{1}{T}\mathbb{E}\left[\log\det\left(\mathbf{I}_{N}+\frac{P_{femtocell}}{N\sigma_{n}^{2}}\gamma_{f_{kl}}\mathbf{h}_{f_{kl}}\mathbf{h}_{f_{kl}}^{T}\right),\right]
\end{equation}

where $f_{kl}\in G_{l}$.

\section{Blind Interference Alignment}

\uline{In the macrocell}, The \emph{$NT\times1$} signal at receiver
$a_{k}$, for the supersymbol, is given by (\ref{eq:MacrocellSignal}). 

The total transmit power, as initially presented in \cite{Paper1},
is given by the power constraint: 

\begin{equation}
P_{\mathrm{macrocell}}=\mathbb{E}[\mathrm{tr}(\mathbf{x}_{A}\mathbf{x}_{A}^{T})]=KNa^{2}.\label{eq:59-1}
\end{equation}

Then, the $NT\times1$ transmitted vector $\mathbf{x}_{A}$ is given
by:

\begin{equation}
\mathbf{x}_{A}=\sum_{i=1}^{K}\mathbf{V}^{[a_{i}]}\mathbf{u}^{[a_{i}]},\label{eq:lloo}
\end{equation}

where $\mathbf{u}^{[a_{k}]}$ is the $N\times1$ data stream vector
of each user\emph{ $a_{k}$,} and $\mathbf{V}^{[a_{k}]}$ is the $NT\times N$
beamforming matrix of user $a_{k}$. As mentioned in \cite{Paper1},
the choice of the $NT\times N$ beamforming matrices $\mathbf{V}^{[a_{k}]}$
carrying messages to users in the macrocell is not unique and should
lie in a space that is orthogonal to the channels of the other $K-1$
macrocell users. The beamforming matrix for user $a_{k}$ is given
by (\cite{Paper1}, (2)):
\begin{equation}
\mathbf{V}^{[a_{k}]}=\frac{a}{\sqrt{N}}(\mathbf{v}^{[a_{k}]}\otimes\mathbf{I}_{N}),
\end{equation}

where $a\in\mathrm{\mathfrak{\boldsymbol{\mathfrak{\mathcal{\mathbb{R}}}}}}$
is a constant determined by power considerations (see (\ref{eq:59-1})),
and $T\times1$ $\mathbf{v}^{[a_{k}]}$ should be a unit vector with
entries equal to $c$, $\sqrt{1-c^{2}}$ (for $c\in\mathrm{\mathfrak{\boldsymbol{\mathfrak{\mathcal{\mathbb{R}}}}}}$
and $c\neq0,\pm1$) or $0$, with a different combination for every
$a_{k}$. For every macrocell user, there will be one time slot in
which only they will be receiving messages. Also, there will be another
time slot (time slot 1 in Figure 2) over which \emph{$T_{xA}$} will
transmit to all users. 

\begin{figure}
\begin{centering}
\includegraphics[width=0.46\columnwidth]{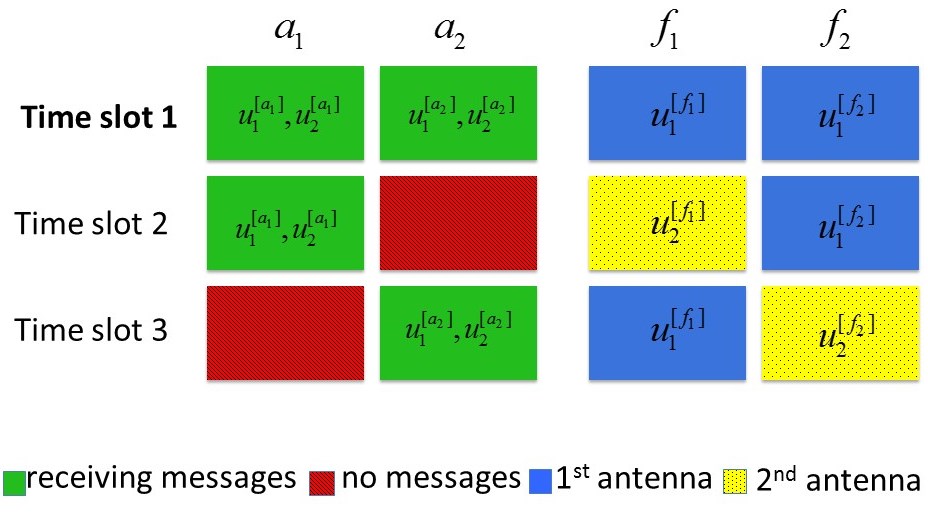}
\par\end{centering}

\caption{Beamforming in the macrocell and femtocells.}
\end{figure}

\begin{example}
The beamforming matrices, as shown in Figure 2, are given by:
\[
\mathbf{V}^{[a_{1}]}=\frac{a}{\sqrt{N}}(\mathbf{v}^{[a_{1}]}\otimes\mathbf{I}_{2})=\frac{a}{\sqrt{2}}\left(\begin{bmatrix}c & \sqrt{1-c^{2}} & 0\end{bmatrix}^{T}\otimes\mathbf{I}_{2}\right),
\]
\[
\mathbf{V}^{[a_{2}]}=\frac{a}{\sqrt{N}}(\mathbf{v}^{[a_{2}]}\otimes\mathbf{I}_{2})=\left(\frac{a}{\sqrt{2}}\begin{bmatrix}c & 0 & \sqrt{1-c^{2}}\end{bmatrix}^{T}\otimes\mathbf{I}_{2}\right).
\]

\end{example}
\uline{At each femtocell}, for Group $G_{1}$, the \emph{$NT\times1$}
signal at receiver $f_{kl}$, for the supersymbol, is given by:
\begin{equation}
\mathbf{y}_{f_{kl}}=\mathbf{H}_{f_{kl}}\mathbf{x}{}_{f_{kl}}+\mathbf{H}_{Af_{kl}}\mathbf{x}_{A}+\mathbf{H}_{f_{k2}f_{kl}}\mathbf{x}_{f_{k2}}+\mathbf{Z}f_{kl},\label{eq:femtoreceived-1}
\end{equation}

and for Group $G_{2}$, the \emph{$NT\times1$} signal at receiver
$f_{k2}$, for the supersymbol, is given by:

\begin{equation}
\mathbf{y}_{f_{k2}}=\mathbf{H}_{f_{k2}}\mathbf{x}{}_{f_{k2}}+\mathbf{H}_{Af_{k2}}\mathbf{x}_{A}+{\displaystyle \sum_{\underset{l\neq2}{l=1,...,L}}}\mathbf{H}_{f_{kl}f_{k2}}\mathbf{x}_{f_{kl}}+\mathbf{Z}_{f_{k2}}.\label{eq:femtoreceived-1-1}
\end{equation}

For Group $G_{1}$, the total transmit power is given by the power
constraint: 

\begin{equation}
P_{femtocell_{1}}=\mathcal{M}_{1}\frac{b_{1}^{2}}{N},\label{eq:Powerfemtos-1-1}
\end{equation}

and the $NT\times1$ vector, transmitted by \emph{$T_{x_{fl}}$} is
given by:

\begin{equation}
\mathbf{x}{}_{f_{kl}}=\mathbf{V}^{[f_{kl}]}\mathbf{U}^{[f_{kl}]},\label{eq:femtovector-1}
\end{equation}
where $\mathbf{U}^{[f_{kl}]}$ is the $\mathcal{M}_{1}\times1$ data
stream vector of each user\emph{ $f_{kl}$}, and $\mathbf{V}^{[f_{kl}]}$
the \emph{$NT\times\mathcal{M}_{1}$} beamforming matrix given by:

\textbf{
\begin{equation}
\mathbf{V}^{[f_{kl}]}=\frac{b_{1}}{\sqrt{N}}\left(\sum_{i=1}^{T-1}\left(\mathbf{\xi}_{i}^{[f_{kl}]^{T}}\otimes\left(\mathbf{r}\mathbf{q}_{i}^{[f_{kl}]}\right)\right)\right),\label{eq:FMETOS}
\end{equation}
}

where $b_{1}\in\mathrm{\mathfrak{\boldsymbol{\mathfrak{\mathcal{\mathbb{R}}}}}}$
is a constant determined by power considerations (see (\ref{eq:Powerfemtos-1-1})),
and $\mathbf{v}^{[f_{kl}]}=\sum_{i=1}^{T-1}\mathbf{\xi}_{i}^{[f_{kl}]}$
is an\emph{ }$1\times T$ vector, and for $i=1,...,T-1$, $\mathbf{\xi}_{i}^{[f_{kl}]}$
has one entry equal to $d$ or $\sqrt{1-(T-2)(N-(L-1))d^{2}}$ (for
$d\in\mathrm{\mathfrak{\boldsymbol{\mathfrak{\mathcal{\mathbb{R}}}}}}$
and $d\neq0,\pm\sqrt{\frac{1}{(T-2)(N-L+1)}}$), and the rest of its
entries equal to $0$, such that $\mathbf{v}^{[f_{kl}]}=\sum_{i=1}^{T-1}\mathbf{\xi}_{i}^{[f_{kl}]}$
has one entry equal to $0$, one entry equal to $\sqrt{1-(T-2)d^{2}}$,
and the rest of its entries equal to $d$. Moreover, we set $\mathbf{r}$
equal to the first $(N-L+1)$ columns of $\mathbf{I}_{N}$ with $\mathbf{e}_{1}$
equal to the sum of the first $(N-1)$ columns of $\mathbf{I}_{N}$.
Furthermore, for $i\neq t_{1}$ ($t_{1}$ denoting the time slot that
\emph{$T_{xA}$} broadcasts to all users in the macrocell), $\mathbf{q}_{i}^{[f_{kl}]}$
is equal to the submatrix of $\mathbf{I}_{\mathcal{M}_{1}}$ consisting
of rows $((N-2)(i-1)+1,(N-2)i),$ and for $i=t_{1}$ $\mathbf{q}_{i}^{[f_{kl}]}$
is equal to any one of $\mathbf{q}_{i}^{[f_{kl}]}$ for $i\neq t_{1}$.
The $t$th component of $\mathbf{\xi}_{i}^{[f_{k}]}$ being 1 means
that in the $kl$th femtocell, the antennas determined by $\mathbf{r}$
are in use at time $t$ and the messages determined by $\mathbf{q}_{i}^{[f_{kl}]}$
are transmitted. 

For Group $G_{2}$, the total transmit power is given by the power
constraint: 
\begin{equation}
P_{femtocell_{2}}=\mathcal{M}_{2}\frac{b_{2}^{2}}{N},\label{eq:Powerfemtos-1-1-1}
\end{equation}

and the \emph{$NT\times\mathcal{M}_{2}$} beamforming matrix $\mathbf{V}^{[f_{kl}]}$
is given by:

\textbf{
\begin{equation}
\mathbf{V}^{[f_{kl}]}=\frac{b_{2}}{\sqrt{N}}\left(\mathbf{v}^{[f_{kl}]^{T}}\otimes\mathbf{e}_{2}\right),
\end{equation}
}

where $b_{2}\in\mathrm{\mathfrak{\boldsymbol{\mathfrak{\mathcal{\mathbb{R}}}}}}$
is a constant determined by power considerations (see (\ref{eq:Powerfemtos-1-1-1})),
and $\mathbf{v}^{[f_{kl}]}$ is an\emph{ }$1\times T$ unit vector
with its $t_{2}$th entry ($t_{2}$ denoting the time slot that $a_{k}$
receives no interference) equal to 1 and the rest of its entries equal
to 0. Vector $\mathbf{e}_{2}$ is equal to the last column of $\mathbf{I}_{N}$.

\subsection{Interference Management}

\uline{In the macrocell}, in order remove inter- and intra-cell
interference, the received signal should be projected to a subspace
orthogonal to the subspace that interference lies in. 
\begin{defn}
The rows of the $N\times NT$ projection matrix $\mathbf{P}^{[a_{k}]}={\displaystyle \sum_{s=1}^{2}}\left(\mathbf{w}_{s}^{[a_{k}]}\otimes\left(\mathbf{D}_{s}^{[a_{k}]}\widetilde{\mathbf{h}}^{[f_{k}a_{k}]}\right)\right)$,
form an orthonormal basis of this subspace, where\end{defn}
\begin{enumerate}
\item for all s, the $1\times T$ $\mathbf{w}_{s}^{[a_{k}]}$ is a unit
vector orthogonal to $\mathbf{v}^{[a_{i}]}$ for $i\neq k$, 
\item \textbf{$\mathbf{w}_{s}^{[a_{k}]}$ }has coefficients equal to zero
on the non-zero values of $\mathbf{\xi}_{i}^{[f_{kl}]^{T}}$ for $i=1,2,...,T-1$,
$s=2$, $l=1,...,L$ and $l\neq2$, and $\mathbf{\mathbf{v}}^{[f_{kl}]^{T}}$for
$s=1$ and $l=2$,
\item \textbf{$\mathbf{D}_{1}^{[a_{k}]}=\mathrm{diag}(\mathbf{e}_{2})$
}and \textbf{$\mathbf{D}_{2}^{[a_{k}]}=\mathrm{diag}(\mathbf{e}_{1})$, }
\item \textbf{$\mathbf{\widetilde{h}}^{[f_{k}a_{k}]}$ }is an $N\times N$
matrix, whose rows are unit vectors, with the $N$th row orthogonal
to all the columns of $\left(\mathbf{h}^{[f_{kl}a_{k}]}\mathbf{r}\right)$
for all $l=1,...,L$ and $l\neq2$, and the remaining $(N-1)$ rows
orthogonal to the columns $\left(\mathbf{h}^{[f_{kl}a_{k}]}\mathbf{\mathbf{e}}_{2}\right)$
for $l=2$.\end{enumerate}
\begin{thm}
Multiplying the received signal by projection matrix $\mathbf{P}^{[a_{k}]}$:
\begin{equation}
\mathbf{\widetilde{y}}^{[a_{k}]}=\mathcal{\mathbf{P}}^{[a_{k}]}\mathbf{y}^{[a_{k}]}=\mathcal{\mathbf{\mathcal{H}}}^{[a_{k}]}\mathbf{U}^{[a_{k}]}+\mathbf{\widetilde{Z}}^{[a_{k}]},\label{eq:Paper1ast-1}
\end{equation}
where 
\begin{equation}
\mathcal{\mathbf{\mathcal{H}}}^{[a_{k}]}=\frac{a}{\sqrt{N}}\mathcal{D}^{[a_{k}]}\mathcal{K}^{[a_{k}]},
\end{equation}
with diagonal matrix 
\begin{equation}
\mathcal{D}^{[a_{k}]}={\displaystyle \sum_{s=1}^{2}}\mathbf{w}_{s}^{[a_{k}]}\mathbf{v}^{[a_{k}]}\mathbf{D}_{s}^{[a_{k}]}=\mathrm{diag}\left(\mathbf{w}_{s}^{[a_{k}]}\mathbf{v}^{[a_{k}]}\right),\label{eq:paper1-13-1}
\end{equation}
and
\begin{equation}
\mathcal{K}^{[a_{k}]}=\sqrt{\gamma_{a_{k}}}\mathbf{\widetilde{h}}^{[f_{k}a_{k}]}\mathbf{h}^{[a_{k}]},\label{eq:paper1-13-1-1}
\end{equation}
and $\mathbf{\widetilde{Z}}^{[a_{k}]}=\mathcal{\mathbf{P}}^{[a_{k}]}\mathbf{Z}^{[a_{k}]}$
remains white noise with the same variance (since $\mathbf{w}_{s}^{[a_{k}]}$
is a unit vector).
\end{thm}
In the femtocells and for Group $G_{1}$, in order remove inter- and
intra-cell interference, the received signal should be projected to
a subspace orthogonal to the subspace that interference lies in. 
\begin{defn}
For Group $G_{1}$, the rows of the $\mathcal{M}_{1}\times NT$ projection
matrix $\mathbf{P}^{[f_{kl}]}=\mathbf{w}\otimes\mathbf{W}^{[f_{kl}]}$
, form an orthonormal basis of this subspace, where\emph{$ $}\end{defn}
\begin{enumerate}
\item the $1\times T$ $\mathbf{w}$ is a unit vector orthogonal to $\mathbf{v}^{[a_{i}]}$
for all \emph{$i$,}
\item and $\mathbf{W}^{[f_{kl}]}$ is an $\mathcal{M}_{1}\times N$ matrix
whose rows are orthogonal to the columns of $\left(\mathbf{h}^{[f_{k2}f_{kl}]}\mathbf{e}_{2}\right)$.\end{enumerate}
\begin{thm}
Multiplying the received signal by projection matrix $\mathbf{P}^{[f_{kl}]}$:
\begin{equation}
\mathbf{\widetilde{y}}^{[f_{kl}]}=\mathcal{\mathbf{P}}^{[f_{kl}]}\mathbf{y}^{[f_{kl}]}=\mathcal{\mathbf{\mathcal{H}}}^{[f_{kl}]}\mathbf{U}^{[f_{kl}]}+\mathbf{\widetilde{Z}}^{[f_{kl}]},\label{eq:last-1}
\end{equation}
where the $\mathcal{M}_{1}\times\mathcal{M}_{1}$ effective channel
matrix is given by: \emph{
\begin{equation}
\mathcal{\mathbf{\mathcal{H}}}^{[f_{kl}]}=\frac{b_{1}}{\sqrt{N}}\mathcal{K}^{[f_{kl}]}\mathcal{D}^{[f_{kl}]},
\end{equation}
}where \emph{
\begin{equation}
\mathcal{K}^{[f_{kl}]}=\sqrt{\gamma_{f_{kl}}}\mathbf{W}^{[f_{kl}]}\mathbf{h}^{[f_{kl}]}\mathbf{r},\;\mathcal{D}^{[f_{kl}]}={\displaystyle \sum_{s=1}^{T-1}}\mathbf{w}\xi_{i}^{[f_{kl}]^{T}}\mathbf{q}_{i}^{[f_{kl}]},
\end{equation}
}and $\mathbf{\widetilde{Z}}^{[f_{kl}]}=\mathcal{\mathbf{P}}^{[f_{kl}]}\mathbf{Z}^{[f_{kl}]}$
remains white noise with the same variance (since $\mathbf{w}$ is
a unit vector).\end{thm}
\begin{defn}
For Group $G_{2}$, the rows of the $\mathcal{M}_{2}\times NT$ projection
matrix $\mathbf{P}^{[f_{kl}]}=\mathbf{w}\otimes\mathbf{W}^{[f_{kl}]},$
, form an orthonormal basis of this subspace, where\end{defn}
\begin{enumerate}
\item the $1\times T$ $\mathbf{w}$ is a unit vector orthogonal to $\mathbf{v}^{[a_{i}]}$
for all \emph{$i$,}
\item and $\mathbf{W}^{[f_{kl}]}$ is an $\mathcal{M}_{2}\times N$ vector
orthogonal to the columns of $\left(\mathbf{h}^{[f_{kl}f_{k2}]}\mathbf{r}\right)$
for $l=1,...,L$ and $l\neq2$.\end{enumerate}
\begin{thm}
Multiplying the received signal by projection matrix $\mathbf{P}^{[f_{kl}]}$:
\emph{
\begin{equation}
\mathbf{\widetilde{y}}^{[f_{kl}]}=\mathcal{\mathbf{P}}^{[f_{kl}]}\mathbf{y}^{[f_{kl}]}=\mathcal{\mathbf{\mathcal{H}}}^{[f_{kl}]}\mathbf{U}^{[f_{kl}]}+\mathbf{\widetilde{Z}}^{[f_{kl}]},,\label{eq:last-1-1}
\end{equation}
}where the $\mathcal{M}_{2}\times\mathcal{M}_{2}$ effective channel
matrix (actually a number), is given by: 
\begin{equation}
\mathcal{\mathbf{\mathcal{H}}}^{[f_{kl}]}=\frac{b_{B}}{\sqrt{N}}\mathcal{\delta}^{[f_{kl}]}\mathcal{K}^{[f_{kl}]},
\end{equation}
where
\begin{equation}
\mathcal{\delta}^{[f_{kl}]}=\mathbf{w}\mathbf{v}^{[f_{kl}]^{T}},\;\mathcal{K}^{[f_{kl}]}=\sqrt{\gamma_{f_{kl}}}\mathbf{W}^{[f_{kl}]}\mathbf{h}^{[f_{kl}]}\mathbf{e}_{2}
\end{equation}
and $\mathbf{\widetilde{Z}}^{[f_{kl}]}=\mathcal{\mathbf{P}}^{[f_{kl}]}\mathbf{Z}^{[f_{kl}]}$
remains white noise with the same variance (since $\mathbf{w}$ is
a unit vector).
\end{thm}

\subsection{Achievable Sum Rate}

In the macrocell, since there is no CSIT, the total rate for each
user for ONE time slot, is given by:

\textbf{
\begin{equation}
R^{[a_{k}]}=\frac{1}{T}\mathrm{\mathtt{\mathbb{E}}}\left[\log\det\left(\mathbf{I}_{N}+\frac{P_{\mathrm{macrocell}}}{KN^{2}\sigma_{n}^{2}}\mathcal{D}^{[a_{k}]}\mathcal{K}^{[a_{k}]}\mathcal{K}^{[a_{k}]^{*}}\mathcal{D}^{[a_{k}]^{*}}\right)\right].\label{eq:RatemacroBIA}
\end{equation}
}

For any channel realization, in the high SNR limit, the rate is maximized
by maximizing the value of 

\begin{equation}
\mathrm{det}\mathcal{D}^{[a_{k}]}=\prod_{s=1}^{2}\left(\mathbf{w}_{s}^{[a_{k}]}\mathbf{v}^{[a_{k}]^{T}}\right).\label{eq:det1-1-1}
\end{equation}

For Group $G_{1}$, in each femtocell, since there is no CSIT, the
rate for each user, for ONE time slot, is given by:

\begin{equation}
R^{[f_{kl}]}=\frac{1}{T}\mathcal{\mathbb{E}}\left[\log\det\left(\mathbf{I}_{\mathcal{M}_{1}}+\frac{P_{\mathrm{femtocell}}}{\mathcal{M}_{1}\sigma_{n}^{2}}\mathcal{K}^{[f_{kl}]}\mathcal{D}^{[f_{kl}]}\mathcal{D}^{[f_{kl}]^{*}}\mathcal{K}^{[f_{kl}]^{*}}\right)\right].
\end{equation}

For Group $G_{2}$, in each femtocell, since there is no CSIT, the
rate for each user, for ONE time slot, is given by:

\begin{equation}
R^{[f_{kl}]}=\frac{1}{T}\mathcal{\mathbb{E}}\left[\log\det\left(1+\frac{P_{\mathrm{femtocell}}}{\sigma_{n}^{2}}\mathcal{\delta}^{[f_{kl}]}\mathcal{K}^{[f_{kl}]}\mathcal{K}^{[f_{kl}]^{*}}\right)\right].
\end{equation}

\section{Special Case of Blind Interference Alignment: $L=1$}

\begin{figure}
\begin{centering}
\begin{minipage}[t]{0.45\textwidth}%
\begin{center}
\includegraphics[width=6cm,height=5cm]{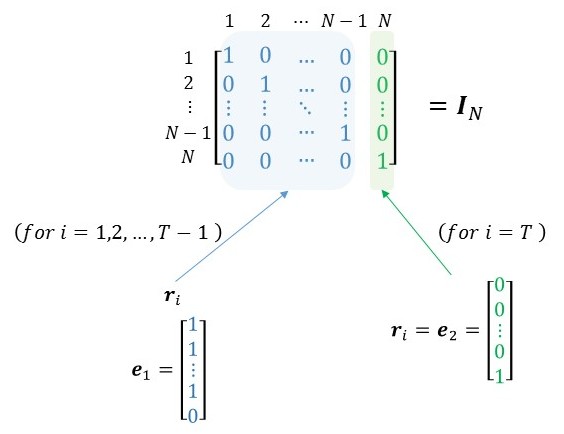}
\par\end{center}%
\end{minipage}\thinspace %
\begin{minipage}[t]{0.45\textwidth}%
\begin{center}
\includegraphics[width=6cm,height=5cm]{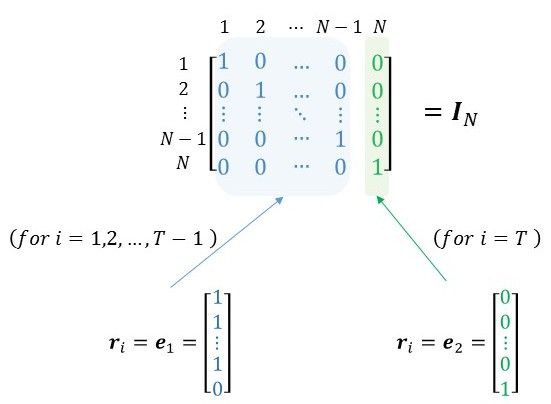}
\par\end{center}%
\end{minipage}
\par\end{centering}

\caption{Design of $r_{i}$ and $e_{i}$: (left) for $M_{r}=N$, and (right)
for r $M_{r}=N-1$. \label{fig:Journal1-1-1-2-1-1} }
\end{figure}

For the special case of $L=1$ (i.e. only one femtocell interfering
with every user in the macrocell), which is the case considered in
this paper, only one group $G_{1}$ exists, and the $M_{r}$ receive
antennas in the femtocells can be equal to $N$ or $N-1$. Furthermore,
the \emph{$NT\times\mathcal{M}_{1}$} beamforming matrix $\mathbf{V}^{[f_{k}]}$
is given by:

\textbf{
\begin{equation}
\mathbf{V}^{[f_{k}]}=\frac{b_{1}}{\sqrt{N}}\left(\sum_{i=1}^{T}\mathbf{\mathbf{\xi}}_{i}^{[f_{k}]^{T}}\otimes\mathbf{r}_{i}\mathbf{q}_{i}\right),
\end{equation}
}

where $\mathbf{v}_{1}^{[f_{k}]}=\sum_{i=1}^{T-1}\mathbf{\xi}_{i}^{[f_{k}]}$
and $\mathbf{v}_{2}^{[f_{k}]}=\mathbf{\xi}_{T}^{[f_{k}]}$ are\emph{
}$1\times T$ vectors. For $i=1,...,T-1$, $\mathbf{\xi}_{i}^{[f_{k}]}$
has one entry equal to $d$ (for $d\in\mathrm{\mathfrak{\boldsymbol{\mathfrak{\mathcal{\mathbb{R}}}}}}$
and $d\neq0,\pm\sqrt{\frac{1}{(T-1)(M_{r}-1)}}$), and $T-1$ entries
equal to $0$, such that $\mathbf{v}_{1}^{[f_{k}]}=\sum_{i=1}^{T-1}\mathbf{\xi}_{i}^{[f_{k}]}$
has $T-1$ entries equal to $d$ and one entry equal to $0$. Vector
$\mathbf{v}_{2}^{[f_{k}]}$ has only its $t_{1}$th entry ($t_{1}$
denoting the time slot that $a_{k}$ receives no interference) equal
to $\sqrt{1-(T-1)(M_{r}-1)d^{2}}$ and all the rest equal to 0, such
that $\sum_{j=1}^{2}\mathbf{v}_{j}^{[f_{k}]}$ has no zero elements
for every $k$. 
\begin{itemize}
\item If $M_{r}=N$, for $i=1,...,T-1$, we set $\mathbf{r}_{i}$ equal
to the first $N-1$ columns of $\mathbf{I}_{N}$ with $\mathbf{e}_{1}$
equal to the sum of the columns of $\mathbf{r}_{i}$, and $\mathbf{e}_{2}=\mathbf{r}_{T}$
equal to the last column of $\mathbf{I}_{N}$ (see Figure \ref{fig:Journal1-1-1-2-1-1}
(left)). 
\item If $M_{r}=N-1$, for $i=1,...,T-1$, we set $\mathbf{r}_{i}$ equal
to the sum of the first $M_{r}$ columns of $\mathbf{I}_{N}$ with
$\mathbf{\mathbf{e}}_{\mathbf{1}}=\mathbf{r_{i}}$ for any $i$, and
$\mathbf{e}_{2}=\mathbf{r}_{T}$ equal to the last column of $\mathbf{I}_{N}$
(see Figure \ref{fig:Journal1-1-1-2-1-1} (right)). 
\end{itemize}

Furthermore, for $i=1,...,T-1$ and $i\neq t_{2}$ ($t_{2}$ denoting
the time slot that \emph{$T_{xA}$} broadcasts to all users in the
macrocell), $\mathbf{q}_{i}$ is equal to the submatrix of $\mathbf{I}_{\mathcal{M}_{1}}$
consisting of rows $(M_{r}(i-1)+1,M_{r}i),$ $\mathbf{q}_{T}$ is
equal to the submatrix of $\mathbf{I}_{\mathcal{M}}$ consisting of
row $\mathcal{M}_{1}$, and $\mathbf{q}_{t_{2}}$ is equal to any
one of $\mathbf{q}_{i}$ for $i=1,...,T-1$ and $i\neq t_{2}$. The
$t$th component of $\mathbf{\xi}_{i}^{[f_{k}]}$ being 1 means that
in the $k$th femtocell, the antennas determined by $\mathbf{r}_{i}$
are in use at time $t$, and the messages determined by $\mathbf{q}_{i}$
are transmitted. 
\begin{example}
The beamforming matrix for user \emph{$f_{1}$}, as depicted in Figure
2, is given by:
\[
\mathbf{V}_{f_{1}}=\frac{b}{\sqrt{2}}\left(\sum_{i=1}^{3}\mathbf{\mathbf{\xi}}_{i}^{[f_{1}]^{T}}\otimes\mathbf{r}_{i}\mathbf{q}_{i}\right),
\]
with 
\[
\sum_{i=1}^{2}\mathbf{\xi}_{i}^{[f_{1}]}=\mathbf{v}_{1}^{[f_{1}]}=\begin{bmatrix}d & 0 & d\end{bmatrix},
\]
\[
\xi_{3}^{[f_{1}]}=\mathbf{v}_{2}^{[f_{1}]}=\begin{bmatrix}0 & \sqrt{1-2d^{2}} & 0\end{bmatrix}
\]
For $i=1,2:$
\[
\mathbf{r}_{i}=\mathbf{e}_{1}=\begin{bmatrix}1 & 0\end{bmatrix}^{T},\;\mathbf{r}_{3}=\mathbf{e}_{2}=\begin{bmatrix}0 & 1\end{bmatrix}^{T},
\]
and $\mathbf{q}_{i}$ the \emph{i}th unit basis vector where
\[
\mathbf{q}_{1}=\mathbf{q}_{2}=\begin{bmatrix}1 & 0\end{bmatrix},\;\mathbf{q}_{3}=\begin{bmatrix}0 & 1\end{bmatrix}.
\]

\end{example}

\subsection{Interference Management}

In the macrocell projection matrix $\mathbf{P}^{[a_{k}]}$ is given
by (40). In each femtocell, in order remove inter-cell interference,
the received signal should be projected to a subspace orthogonal to
the subspace that inter-cell interference lies in. 
\begin{example}
For the example model, setting $A=\sqrt{\left(h_{22}^{[f_{1}a_{1}]}\right)^{2}+\left(h_{12}^{[f_{1}a_{1}]}\right)}$
, $B=\sqrt{\left(h_{21}^{[f_{1}a_{1}]}\right)^{2}+\left(h_{11}^{[f_{1}a_{1}]}\right)}$,
$\mathbf{P}^{[a_{1}]}$ is given by:\textbf{
\[
\mathbf{P}^{[a_{1}]}={\displaystyle \sum_{s=1}^{2}}\left(\mathbf{w}_{s}^{[a_{k}]}\otimes\left(\mathbf{D}_{s}^{[a_{k}]}\widetilde{\mathbf{h}}^{[f_{k}a_{k}]}\right)\right)
\]
} where 
\[
\mathbf{w}_{1}^{[a_{1}]}=\begin{bmatrix}-\sqrt{1-c^{2}} & 0 & c\end{bmatrix},\;\mathbf{w}_{2}^{[a_{1}]}=\begin{bmatrix}0 & 1 & 0\end{bmatrix},
\]
\begin{equation}
\mathbf{D}_{1}^{[a_{1}]}=\mathrm{diag}(\begin{bmatrix}0 & 1\end{bmatrix}^{T}),\;\mathbf{D}_{2}^{[a_{1}]}=\mathrm{diag}(\begin{bmatrix}1 & 0\end{bmatrix}^{T}),
\end{equation}
\begin{equation}
\widetilde{\mathbf{h}}^{[f_{1}a_{1}]}=\begin{bmatrix}\frac{-h_{22}^{[f_{1}a_{1}]}}{A} & \frac{h_{12}^{[f_{1}a_{1}]}}{A}\\
\frac{-h_{21}^{[f_{1}a_{1}]}}{B} & \frac{h_{11}^{[f_{1}a_{1}]}}{B}
\end{bmatrix}.
\end{equation}
\end{example}
\begin{defn}
The rows of the $\mathcal{M}_{1}\times M_{r}T$ projection matrix
$\mathbf{P}^{[f_{k}]}$, which is the same for all femtocell users
$f_{k}$, form an orthonormal basis of this subspace:
\begin{equation}
\mathbf{P}^{[f_{k}]}=\mathbf{w}\otimes\mathbf{W},
\end{equation}
where\end{defn}
\begin{enumerate}
\item the $1\times T$ $\mathbf{w}$ is a unit vector orthogonal to $\mathbf{v}^{[a_{i}]}$
for all \emph{$i$,}
\item and $\mathbf{W}$ is an $\mathcal{M}_{1}\times M_{r}$ all-ones matrix.\end{enumerate}
\begin{example}
For the toy-model, $P^{[f_{1}]}$ is given by:\textbf{
\[
\mathbf{P}^{[f_{1}]}=\mathbf{w}\otimes\mathbf{W}=\left[\frac{1}{\sqrt{1+2c^{2}}}\begin{bmatrix}c & -\sqrt{1-c^{2}} & c & c\end{bmatrix}\otimes\left(\begin{bmatrix}1 & 1 & 1\\
1 & 1 & 1
\end{bmatrix}^{T}\right)\right],
\]
}where 
\[
\mathbf{w}=\frac{1}{\sqrt{1+2c^{2}}}\begin{bmatrix}c & -\sqrt{1-c^{2}} & c & c\end{bmatrix},\;\mathbf{W}=\begin{bmatrix}1 & 1 & 1\\
1 & 1 & 1
\end{bmatrix}^{T}.
\]
\end{example}
\begin{thm}
Multiplying the received signal by projection matrix $\mathbf{P}^{[f_{k}]}$:
\begin{equation}
\mathbf{\widetilde{y}}^{[f_{k}]}=\mathcal{\mathbf{P}}^{[f_{k}]}\mathbf{y}^{[f_{k}]}=\mathcal{\mathbf{\mathcal{H}}}^{[f_{k}]}\mathbf{U}^{[f_{k}]}+\mathbf{\widetilde{Z}}^{[f_{k}]},\label{eq:last}
\end{equation}
where the $\mathcal{M}_{1}\times\mathcal{M}_{1}$ effective channel
matrix is given by: 
\begin{equation}
\mathcal{\mathbf{\mathcal{H}}}^{[f_{k}]}=\frac{b}{\sqrt{N}}\mathcal{K}^{[f_{k}]}\mathcal{D}^{[f_{k}]},
\end{equation}
where\emph{ 
\begin{equation}
\mathcal{K}^{[f_{k}]}=\sqrt{\gamma_{f_{k}}}\mathbf{W}\mathbf{h}^{[f_{k}]},\;\mathcal{D}^{[f_{k}]}={\displaystyle \sum_{i=1}^{T}}\mathbf{w}\xi_{i}^{[f_{k}]^{T}}\mathbf{r}_{i}\mathbf{q}_{i},
\end{equation}
}and $\mathbf{\widetilde{Z}}^{[f_{k}]}=\mathcal{\mathbf{P}}^{[f_{k}]}\mathbf{Z}^{[f_{k}]}$
remains white noise with the same variance (since $\mathbf{w}$ is
a unit vector).\end{thm}
\begin{IEEEproof}
We show that $P^{[f_{k}]}$ removes inter-cell interference at the
$k$th receiver. Substituting, (\ref{eq:femtoreceived-1}) and (\ref{eq:femtovector-1})
in (\ref{eq:last}), and using $\left(A\otimes B\right)\left(C\otimes D\right)=\left(AC\right)\otimes\left(BD\right)$,
the coefficient of $\mathbf{U}^{[a_{i}]}$, for all $i$, becomes:
\begin{align}
\mathbf{P}^{[f_{k}]}\mathbf{H}^{[Af_{k}]}\mathbf{V}^{[a_{i}]} & =\frac{a}{\sqrt{N}}\left(\mathbf{w}\otimes\mathbf{W}\right)\sqrt{\gamma_{a_{1}}}\left(\mathbf{I}_{T}\otimes\mathbf{h}^{[a_{i}]}\right)\left(\mathbf{v}^{[a_{i}]}\otimes\mathbf{I}_{N}\right)\nonumber \\
 & =\frac{a}{\sqrt{N}}\sqrt{\gamma_{a_{1}}}\left(\mathbf{w}\mathbf{v}^{[a_{i}]}\right)\otimes\left(\mathbf{W}\mathbf{h}^{[a_{i}]}\right),
\end{align}
where by 1) in Definition 16, for all i, $\mathbf{w}\mathbf{v}^{[a_{i}]}=0$,
i.e. $\mathbf{w}$ is orthogonal to $\mathbf{v}^{[a_{i}]}$\emph{
}for all\emph{ }$i$. 
\end{IEEEproof}

\subsection{Achievable Sum Rate}

In the macrocell, the total rate for each user is given by (\ref{eq:RatemacroBIA}).
For any channel realization, in the high SNR limit, the rate is maximized
for $c=\pm-1/\sqrt{3},$ by maximizing (\ref{eq:det1-1-1}).

In each femtocell, since there is no CSIT, the rate for each user,
for ONE time slot, is given by:

\begin{equation}
R^{[f_{k}]}=\frac{1}{T}\mathcal{\mathbb{E}}\left[\log\det\left(\mathbf{I}_{\mathcal{M}}+\frac{P_{\mathrm{femtocell}}}{\mathcal{M}\sigma_{n}^{2}}\mathcal{K}^{[f_{k}]}\mathcal{D}^{[f_{k}]}\mathcal{D}^{[f_{k}]^{*}}\mathcal{K}^{[f_{k}]^{*}}\right)\right],
\end{equation}

where by taking $\mathrm{det}(\mathcal{D}^{[f_{k}]}\mathcal{D}^{[f_{k}]*})$,
the optimal value of $d$ was calculated as $d=\pm0.5$.

\section{Performance Results}

Our simulations were based on the example model described and were
performed in \emph{Matlab}.\textbf{ }The statistical model chosen
was i.i.d. Rayleigh and our input symbols were Quadrature Phase Shift
Keying (QPSK) modulated. Maximum-Likelihood (ML) detection was performed
in the end of the decoding stage. The total transmit power in the
macrocell was considered as $40$W and in the femtocells as $5$W
(typical values for transmit power in macrocells for 4G systems),
and therefore $a$ and $b$, constants determined by power considerations
in (\ref{eq:Ch7-femtosa}) and (\ref{eq:Ch7-femtosb}) for the hybrid
scheme, and (\ref{eq:59-1}) and (\ref{eq:Powerfemtos-1-1}) for the
Blind IA scheme respectively, are given by $a=\sqrt{40}$ and $b=\sqrt{5}$.
Moreover, simulations were performed for $100-500$ frames, with each
frame consisting of $6144$ bits.

\subsection{Degrees of Freedom}
\begin{thm}
For the hybrid scheme, in a heterogeneous network, the total DoF achieved
are given by $DoF_{Hybrid}=\frac{KN}{T}(1+L),$ where $K/T$ is the
average number of users per group, showing that the less the number
of groups is, the more DoF are provided.
\end{thm}
\begin{table}
\begin{centering}
\begin{tabular}{|c|c|c|c|}
\hline 
\textbf{Scheme} & \textbf{Macrocell} & \textbf{KL Femtocells} & \textbf{Total Network}\tabularnewline
\hline 
\hline 
\textbf{Hybrid } & $\frac{KN}{T}$ & $\frac{KLN}{T}$ & $\frac{KN}{T}\left(1+L\right)$\tabularnewline
\hline 
\textbf{TDMA} & $\frac{\left(T-x\right)N}{T}$ & $\frac{xKN}{T}$ & $\frac{xN\left(K-1\right)+TN}{T}$\tabularnewline
\hline 
\end{tabular}
\par\end{centering}

\caption{DoF of Hybrid scheme and TDMA. \label{tab:DoF-of-BIA-B-1}}
\end{table}

\begin{figure}
\begin{centering}
\begin{minipage}[t]{0.45\textwidth}%
\begin{center}
\includegraphics[width=6cm,height=5cm]{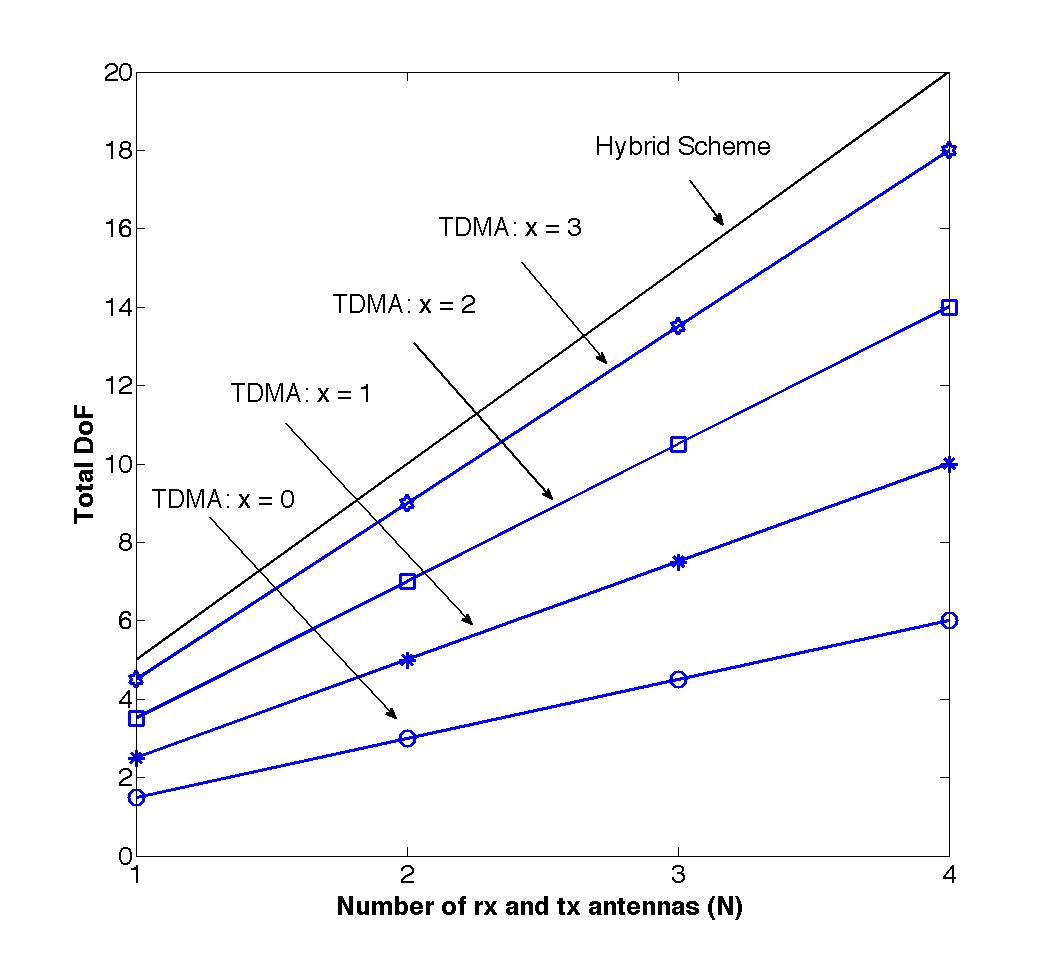}
\par\end{center}%
\end{minipage}\thinspace %
\begin{minipage}[t]{0.45\textwidth}%
\begin{center}
\includegraphics[width=6cm,height=5cm]{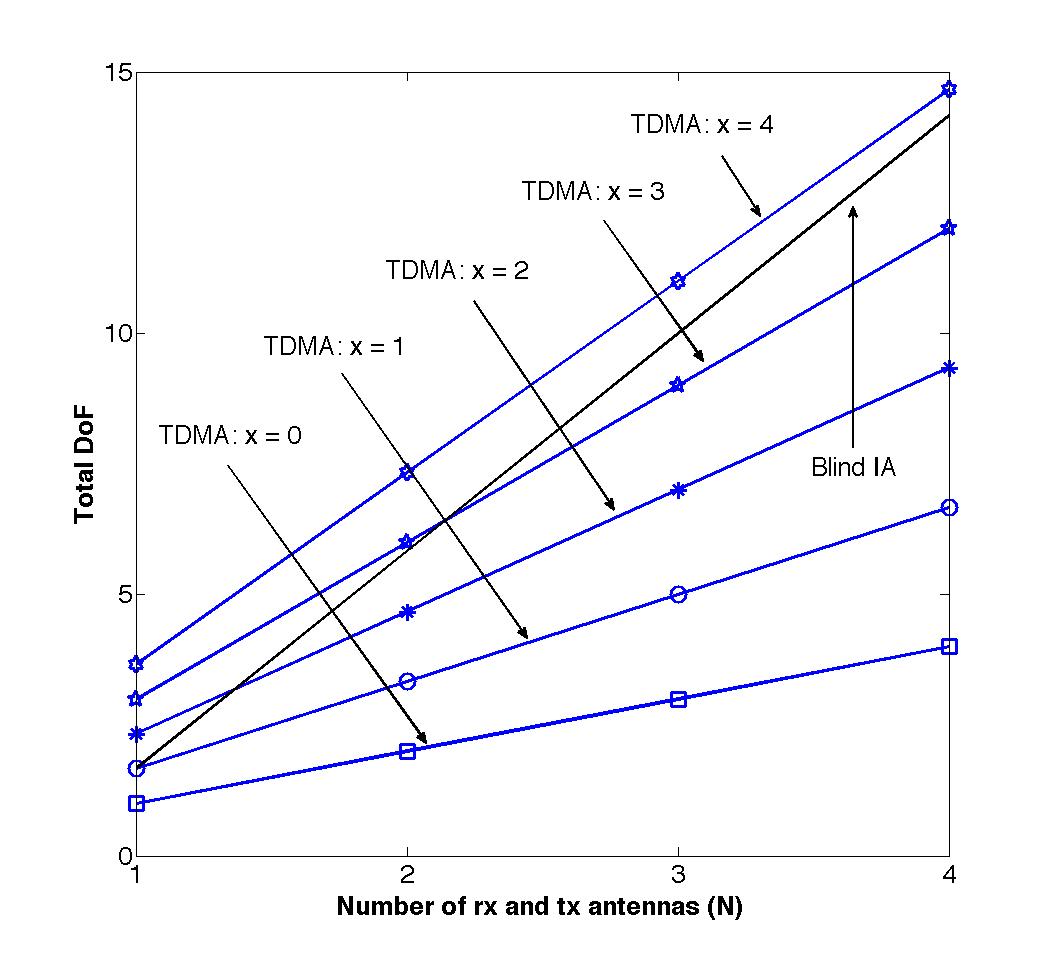}
\par\end{center}%
\end{minipage}
\par\end{centering}

\caption{(left) Hybrid Scheme vs. TDMA - Total DoF for $K=5$, $L=3$, $N=\left\{ 1,2,3,4\right\} $,
$x=\left\{ 0,1,2,3\right\} $ . The hybrid scheme always provides
more total DoF. (right) Blind IA vs. TDMA ($L=1$) - Total DoF for
$K=5$, $N=M_{r}=\left\{ 1,2,3,4\right\} $, $x=\left\{ 0,1,2,3,4\right\} $
. TDMA can outperform Blind IA if more time slots are given to the
femtocells. \label{fig:Journal1-1-1-2-1-1-1-2} }
\end{figure}

\begin{thm}
For TDMA, setting $x\in\mathbb{Z}$ and $x<T$, the total DoF achieved
are given by $DoF_{TDMA}=\frac{xN\left(K-1\right)+TN}{T}$, where
considering a fair time slot allocation to all cells, the DoF will
be a function of $x$ which denotes how many time slots will be given
to each cell.
\end{thm}
Table \ref{tab:DoF-of-BIA-B-1} presents a comparison between the
the hybrid and the TDMA scheme. 
\begin{thm}
In a heterogeneous network, as defined in this paper, the hybrid scheme
outperforms TDMA in terms of total DoF, as shown in Figure \ref{fig:Journal1-1-1-2-1-1-1-2}
(left). The total DoF gain achieved by the hybrid scheme is given
by $DoF_{Hybrid}-DoF_{TDMA}=\frac{N\left(K-1\right)\left(T-x\right)}{T}$.\end{thm}
\begin{IEEEproof}
We show that $DoF_{Hybrid}>DoF_{TDMA}$, using $T=L+1$:
\begin{align}
\frac{KN(1+L)}{T} & >\frac{xN\left(K-1\right)+N\left(K-1\right)}{T}\nonumber \\
NT\left(K-1\right) & >xN\left(K-1\right)\nonumber \\
T & >x,
\end{align}
which is true based on the definition that $x<T$.
\end{IEEEproof}

\begin{thm}
For the Blind IA scheme, in a heterogeneous network, the total DoF
achieved are given by $DoF_{Blind\,IA}=K\frac{(N+(K-1)(L-1)(N-L+1)+1)}{T}$\emph{,
}and for the special case $L=1$ where $M_{r}=\left\{ N-1,N\right\} $
by \emph{$DoF_{Blind\,IA\,L=1}=K\frac{(N+(K-1)(M_{r}-1)+1)}{T}$.}
\end{thm}
\begin{table}[H]
\begin{centering}
\begin{tabular}{|c|c|c|c|c|}
\hline 
 & \textbf{Scheme} & \textbf{Macrocell} & \textbf{K Femtocells} & \textbf{Total Network}\tabularnewline
\hline 
$1<L\leqslant3$ & \textbf{Blind IA} & $\frac{KN}{T}$ & $K\frac{((K-1)(L-1)(N-L+1)+1)}{T}$ & $K\frac{(N+(K-1)(L-1)(N-L+1)+1)}{T}$\tabularnewline
\cline{2-5} 
 & \textbf{TDMA} $x$ odd & $\frac{\left(T-x\right)N}{T}$ &  $\frac{\left(\frac{x-1}{2}\right)KNL+K\left(L-1\right)N}{T}$ & $\frac{N\left(KL+1\right)+\left(\frac{x-1}{2}\right)KNL-xN}{T}$\tabularnewline
\cline{4-5} 
 & $x$ even &  &  $\frac{\frac{x}{2}KNL}{T}$ & $\frac{K\left(N+1\right)+\frac{x}{2}KNL-xN}{T}$\tabularnewline
\hline 
$L=1$ & \textbf{Blind IA} & $\frac{KN}{T}$ & $K\frac{(K-1)(M_{r}-1)+1}{T}$ & $\frac{K(N+(K-1)(M_{r}-1)+1)}{T}$\tabularnewline
\cline{2-5} 
 & \textbf{TDMA} & $\frac{\left(T-x\right)N}{T}$ & $\frac{xKM_{r}}{T}$ & $\frac{x\left(M_{r}K-N\right)+NT}{T}$\tabularnewline
\hline 
\end{tabular}
\par\end{centering}

\caption{DoF of Blind IA and TDMA (Scheme B - $L=1$). \label{tab:DoF-of-BIA-B}}
\end{table}

\begin{thm}
For TDMA, setting $x\in\mathbb{Z}$ and $x<T$, the total DoF achieved,
for $x$ odd and even, are given by $DoF_{TDMA-x\,odd}=\frac{N\left(KL+1\right)+\left(\frac{x-1}{2}\right)KNL-xN}{T}$
and $DoF_{TDMA-x\,even}=\frac{K\left(N+1\right)+\frac{x}{2}KNL-xN}{T}$
respectively, considering a relatively fair time slot allocation to
all cells. For the special case $L=1$ where $M_{r}=\left\{ N-1,N\right\} $,
the DoF are given by $DoF_{TDMA-L=1}=x\left(M_{r}K-N\right)+NT$. 
\end{thm}
Table \ref{tab:DoF-of-BIA-B} presents a comparison between Blind
IA and TDMA. 

\begin{figure}
\begin{centering}
\begin{minipage}[t]{0.45\textwidth}%
\begin{center}
\includegraphics[width=6cm,height=5cm]{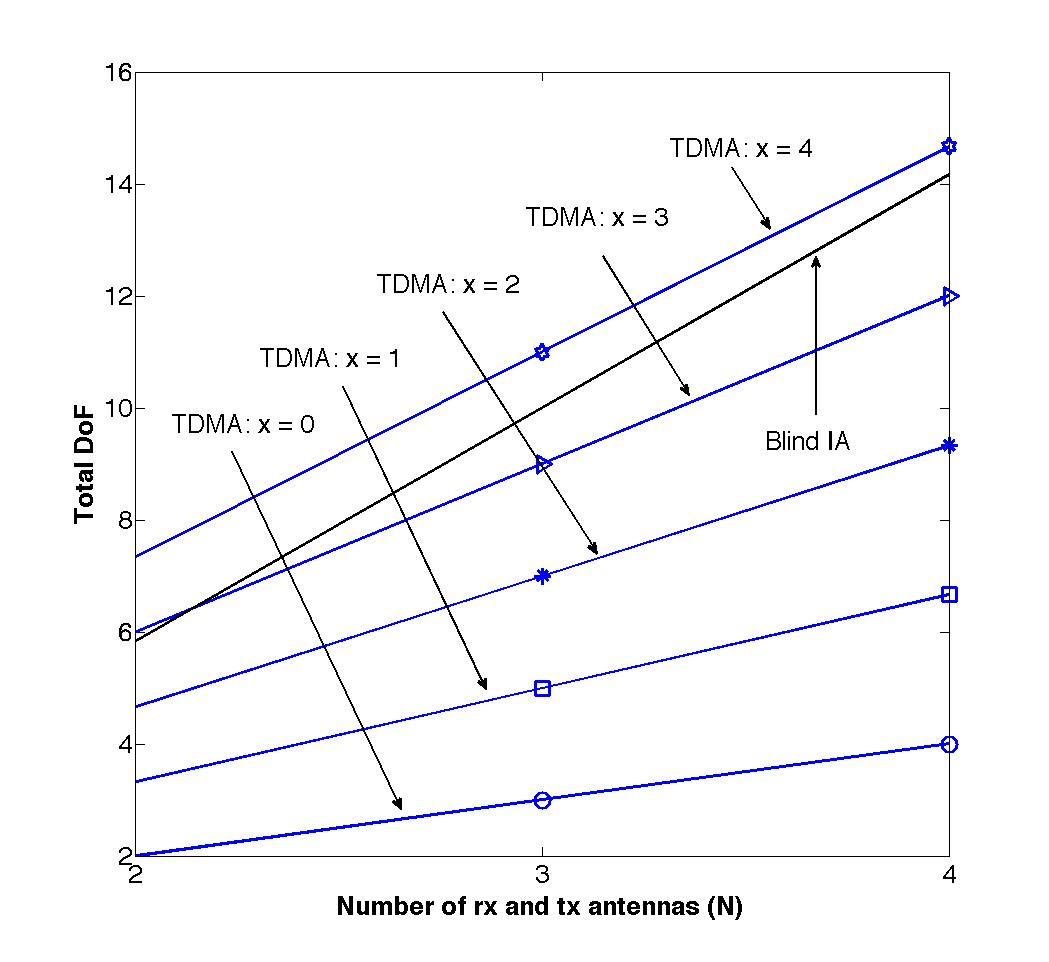}
\par\end{center}%
\end{minipage}\thinspace %
\begin{minipage}[t]{0.45\textwidth}%
\begin{center}
\includegraphics[width=6cm,height=5cm]{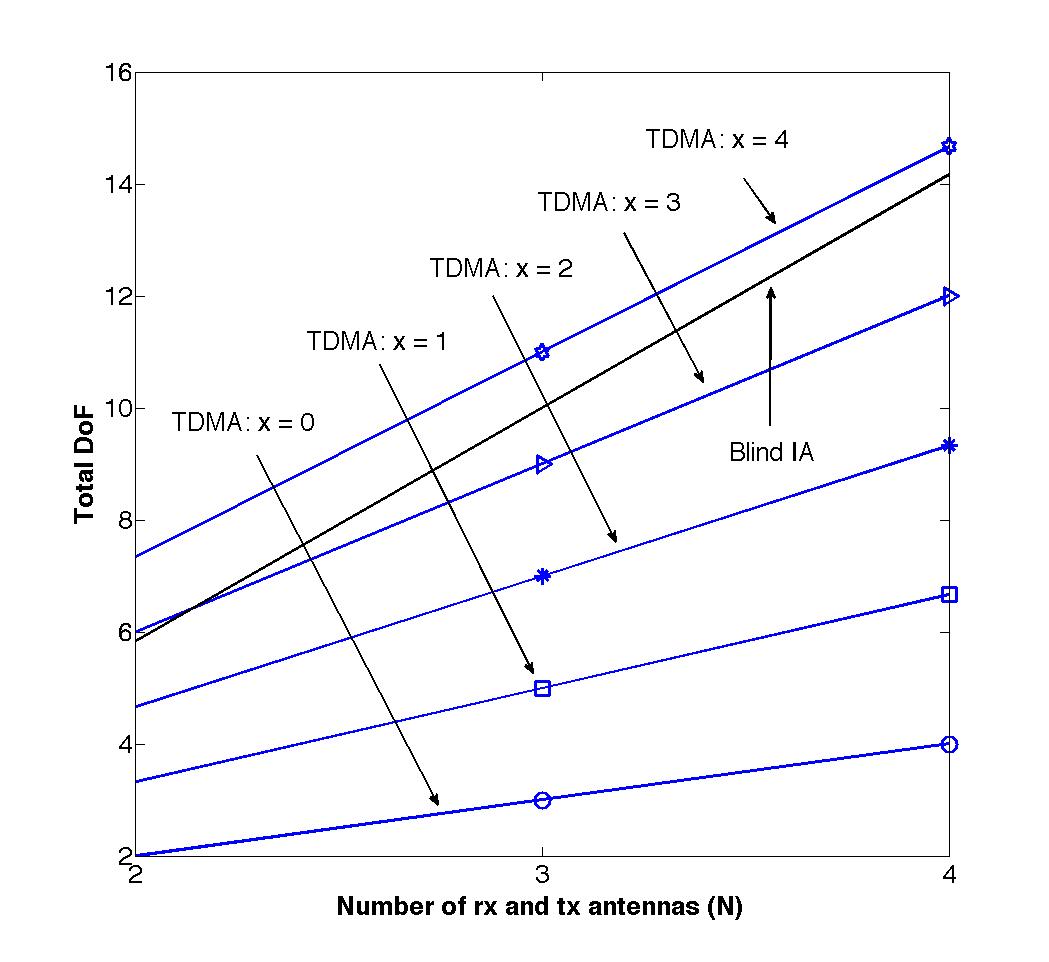}
\par\end{center}%
\end{minipage}
\par\end{centering}

\caption{Blind IA vs. TDMA - Total DoF for $K=5$, $L=2$, $N=\left\{ 2,3,4\right\} $,
$x=\left\{ 0,1,2,3,4\right\} $ for $x$ odd (left) and $x$ even
(right). TDMA outperforms Blind IA as more time slots are dedicated
to femtocells. \label{fig:Journal1-1-1-2-1-1-1} }
\end{figure}

\begin{thm}
For the case that $x$ is odd, the gain of Blind IA to TDMA is given
by $DoF_{Blind\,IA}-DoF_{TDMA-x\,odd}=\frac{2K(N+\mathcal{M}_{1}(L-1)+1)-2N(KL+1)-KNL+xN(KL-2)}{2T}$
only when $\frac{2K(N+\mathcal{M}_{1}(L-1)+1)-2N(KL+1)+KNL}{N(KL-2)}>x$.
For the case that $x$ is even, the gain of Blind IA is given by $DoF_{Blind\,IA}-DoF_{TDMA-x\,even}=\frac{2K(N+(L-1)\mathcal{M}_{1}+1)-2K(N+1)+xN(KL-2)}{2T}$
only when $\frac{2K(N+\mathcal{M}_{1}(L-1)+1)-2K(N+1)}{N(KL-2)}>x$.
For the special case of $L=1$ where $M_{r}=\left\{ N-1,N\right\} $,
the gain of Blind IA is given by $DoF_{Blind\,IA}-DoF_{TDMA-L=1}=\frac{K(N+(K-1)(M_{r}-1)+1)-NT+x\left(N-M_{r}K\right)}{T}$
only when $\frac{K(N+(K-1)(M_{r}-1)+1)-NT}{M_{r}K-N}>x$.
\end{thm}
As shown in Figures \ref{fig:Journal1-1-1-2-1-1-1-2} (right) and
\ref{fig:Journal1-1-1-2-1-1-1}, Blind IA outperforms TDMA in the
case that we provide more time slots to the macrocell, whereas as
the number of time slots dedicated to the femtocells increases, TDMA
can achieve more total DoF. 

\begin{figure}
\begin{centering}
\hfill{}%
\begin{minipage}[t]{0.45\textwidth}%
\begin{center}
\includegraphics[width=6cm,height=5cm]{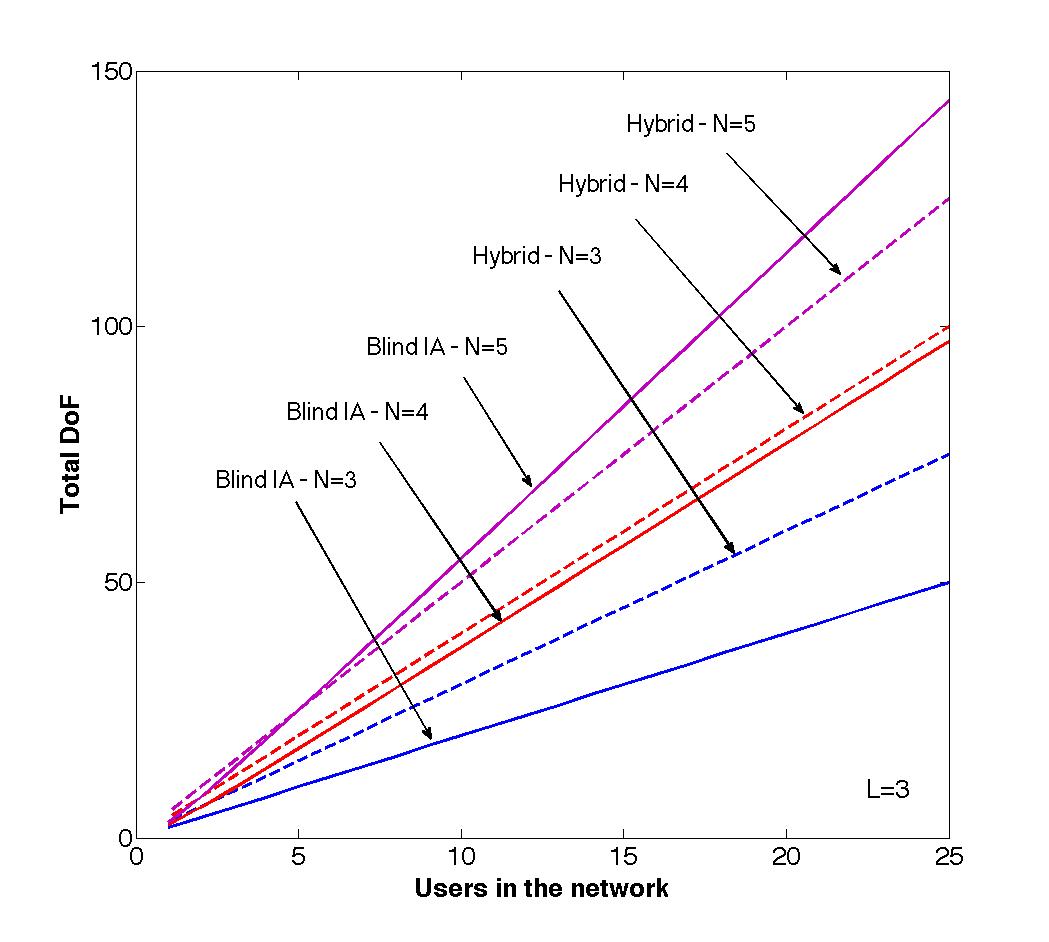}
\par\end{center}%
\end{minipage}\thinspace %
\begin{minipage}[t]{0.45\textwidth}%
\begin{center}
\includegraphics[width=6cm,height=5cm]{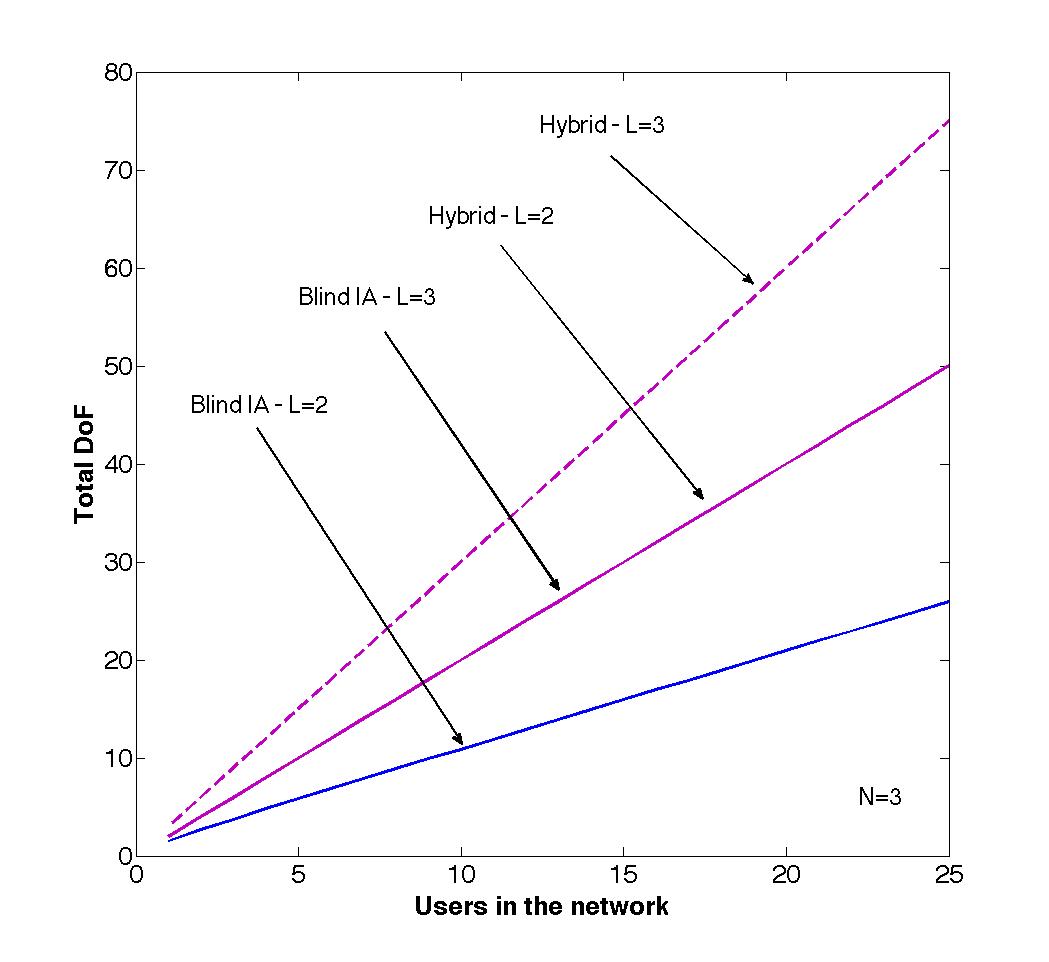}
\par\end{center}%
\end{minipage}
\par\end{centering}

\caption{Hybrid Scheme vs. Blind IA - Total DoF for $N=\left\{ 3,4,5\right\} $,
$L=3$ (left) and $L=\left\{ 2,3\right\} $, $N=3$ (right). \label{fig:Journal1-1-1-2-1} }
\end{figure}

Table \ref{tab:Ch7BIA-hybrid} presents a comparison between the hybrid
scheme and the Blind IA. The hybrid scheme outperforms the Blind IA
mainly due to the fact that less time slots are required to send the
same number of messages. Figure \ref{fig:Journal1-1-1-2-1} (left)
shows that as the number of transmit and receive antennas increases
the benefit we get from the hybrid scheme gets smaller, resulting
in the Blind IA scheme outperforming the hybrid one for $N>4$. Finally,
in Figure \ref{fig:Journal1-1-1-2-1} (right) it can be seen that
as the number of femtocells that interfere with every macrocell user
increases, again the benefit we get from the hybrid schemes decreases.
However, in general the main advantage of the hybrid scheme is that
it overcomes the limitation of the Blind IA scheme that it is valid
only for $L\leqslant3$.

\begin{table}
\begin{centering}
\begin{tabular}{|c|c|c|}
\hline 
\textbf{DoF} & \textbf{Total} & \textbf{Example ($L=1$)}\tabularnewline
\hline 
\hline 
\textbf{Hybrid} & $\frac{KN}{T}\left(1+L\right)$ & $\mathbf{\frac{8}{2}}$\tabularnewline
\hline 
\textbf{Blind IA} & $K\frac{(N+(K-1)(L-1)(N-L+1)+1)}{T}$ & -\tabularnewline
\textbf{Case of $L=1$} & \textbf{$\frac{K(N+(K-1)(M_{r}-1)+1)}{T}$} & $\mathbf{\frac{8}{3}}$\tabularnewline
\hline 
\end{tabular}
\par\end{centering}

\caption{Comparison of the hybrid scheme and Blind IA Scheme B (Special case:
\textbf{$L=1$}). More DoF are provided with the hybrid scheme.\label{tab:Ch7BIA-hybrid}}
\end{table}

\subsection{Bit Error Rate (BER) Performance}

First of all, the BER performance of our example model was investigated.
In general, since we are considering the distance of every user from
the transmitter, users closer to the base station will achieve a better
performance than those at the edge of the cell. 

\begin{figure}
\begin{centering}
\begin{minipage}[t]{0.45\textwidth}%
\begin{center}
\includegraphics[width=6cm,height=5cm]{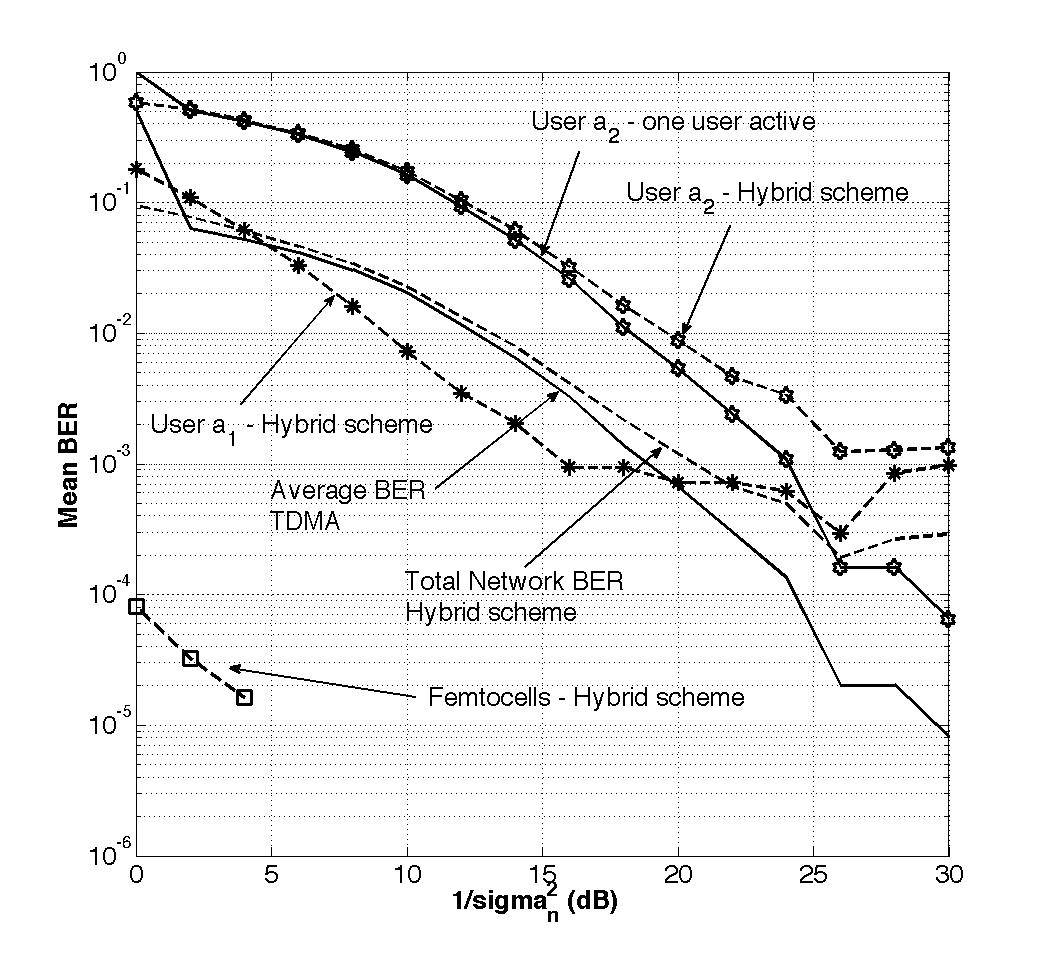}
\par\end{center}%
\end{minipage}\thinspace %
\begin{minipage}[t]{0.45\textwidth}%
\begin{center}
\includegraphics[width=6cm,height=5cm]{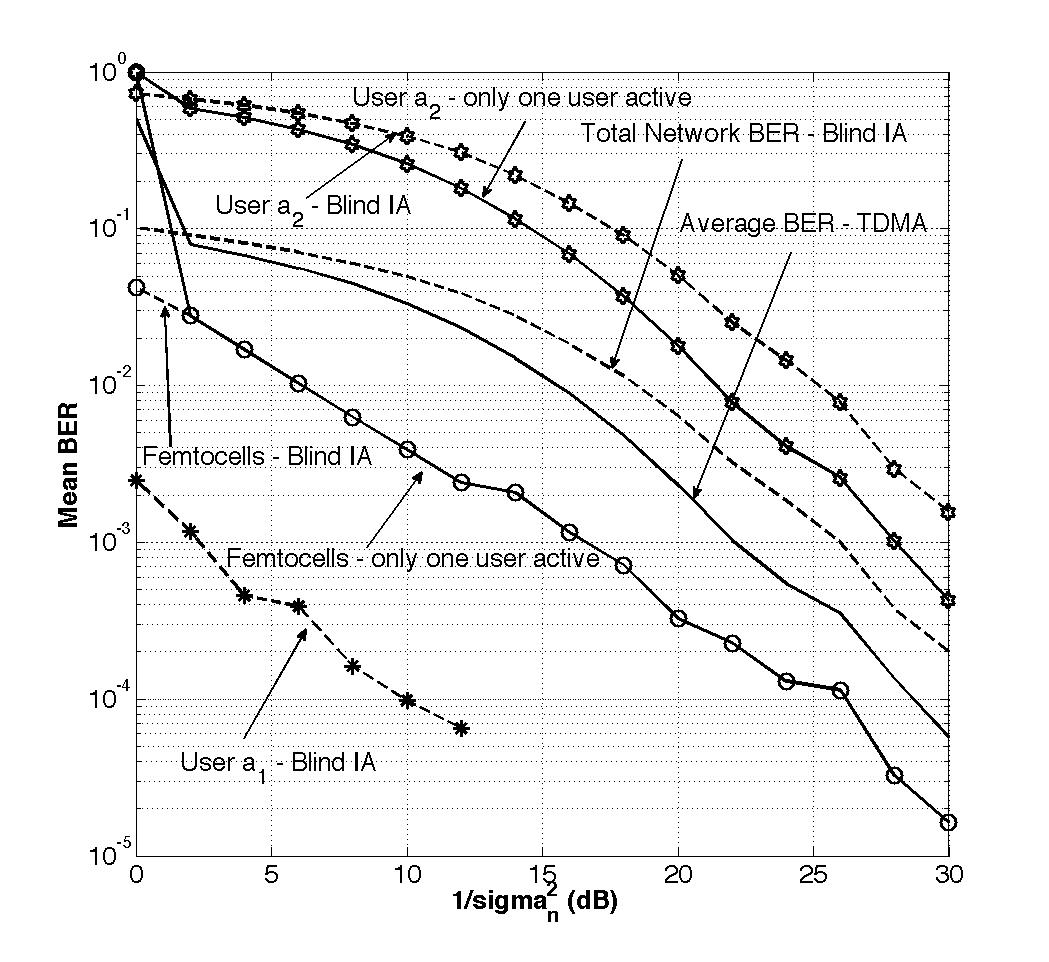}
\par\end{center}%
\end{minipage}
\par\end{centering}

\caption{BER Performance: (left) Hybrid vs. TDMA - Total DoF for $K=5$, $L=2$,
$N=\left\{ 2,3,4\right\} $, $x=\left\{ 0,1,2,3,4\right\} $ for $x$
odd (left) and $x$ even (right) Blind IA vs. TDMA. TDMA outperforms
Blind IA as more time slots are dedicated to femtocells \label{fig:Journal1-1-1-2-1-1-1-1} }
\end{figure}

Both schemes were first compared to the case where only one user is
active in the heterogeneous network (TDMA). Therefore, the BER of
every user, both in the macrocell and femtocells, was simulated assuming
that only them will receive message over $T=2$ and $T=3$ time slots
for the hybrid scheme and Blind IA respectively. Figure \ref{fig:Journal1-1-1-2-1-1-1-1}
(left) depicts the BER for every user separately, for both the hybrid
and the TDMA schemes. Note that the BER for users $a_{1}$, $f_{1}$
and $f_{2}$ is 0 for the range of SNR values. Focusing on the total
network BER for the hybrid scheme and the average BER for the case
of only one user in the network being active, we can observe that
both schemes offer similar overall BER performances, with the average
TDMA BER slightly outperforming the hybrid scheme for high SNR values.
Figure \ref{fig:Journal1-1-1-2-1-1-1-1} (right) depicts the BER for
every user separately, for both the hybrid and the TDMA schemes. Focusing
on the total network BER for the hybrid scheme and the average BER
for the case of only one user in network being active, we can observe
that both schemes offer similar overall BER performances.

\begin{figure}
\begin{centering}
\begin{minipage}[t]{0.45\textwidth}%
\begin{center}
\includegraphics[width=6cm,height=5cm]{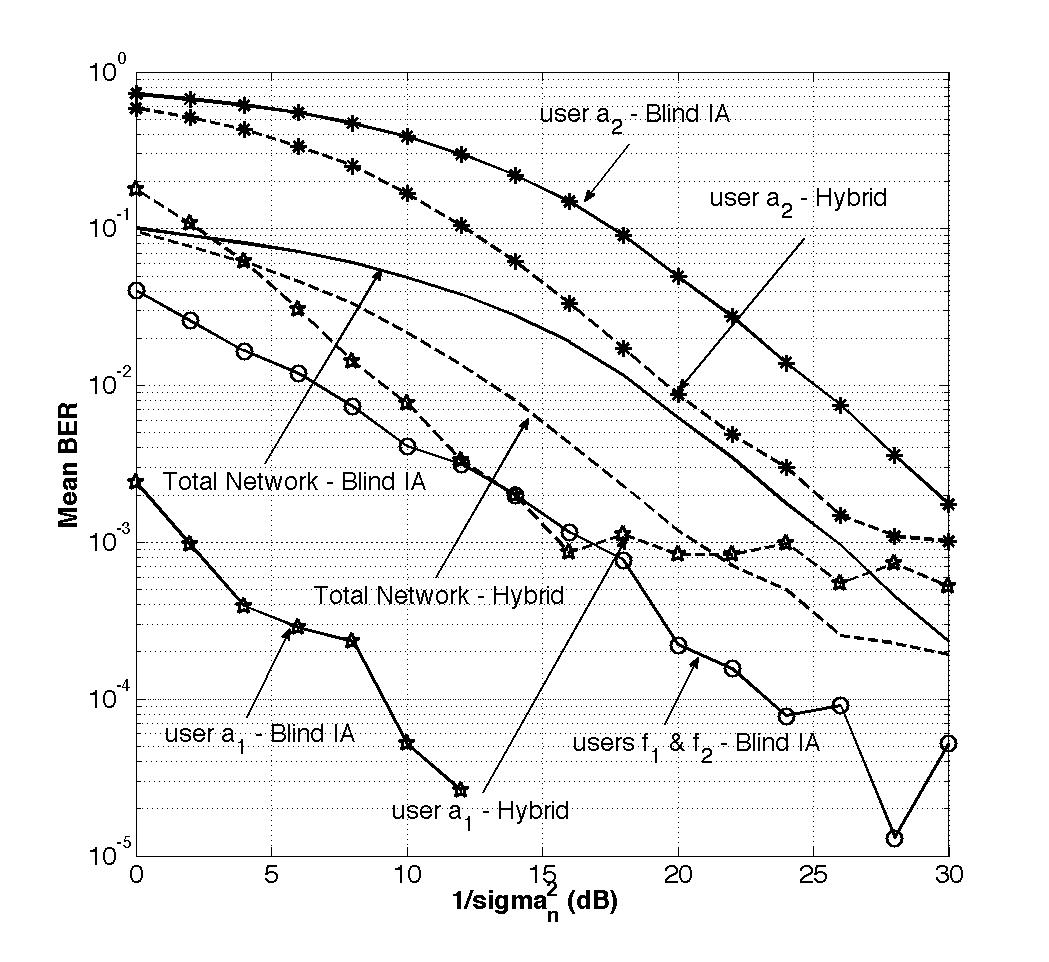}
\par\end{center}%
\end{minipage}\thinspace %
\begin{minipage}[t]{0.45\textwidth}%
\begin{center}
\includegraphics[width=6cm,height=5cm]{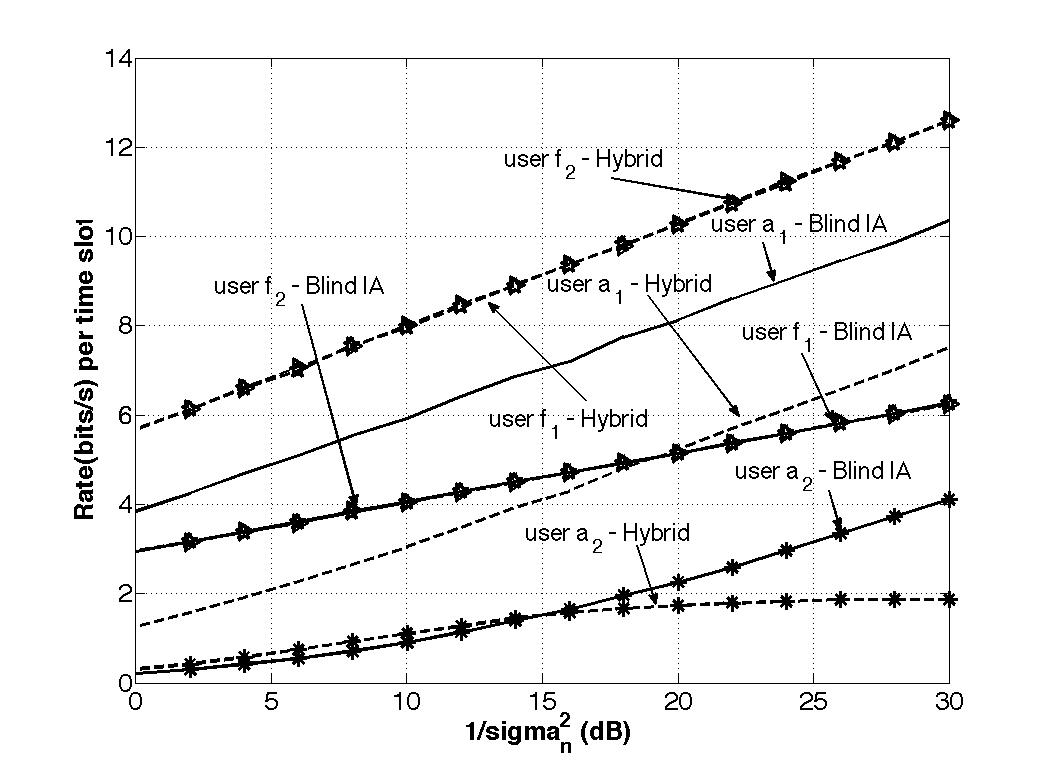}
\par\end{center}%
\end{minipage}
\par\end{centering}

\caption{Hybrid Scheme vs. Blind IA - (left) BER Performance, (right) Sum rate
Performance. Users in the macrocell can achieve better performance
with Blind IA. Users in femtocells achieve better performance with
the hybrid scheme\label{fig:Journal1-1-1-2-1-1-1-1-2} }
\end{figure}

The hybrid scheme was compared to the Blind IA scheme, although a
completely fair comparison is not possible, since Blind IA requires
$T=3$ time slots and the hybrid scheme $T=2$ time slots (the noise
was not the same for the two models), and power allocation is different.
However, path loss and channel gains were considered the same. Figure
\ref{fig:Journal1-1-1-2-1-1-1-1-2} (left) depicts the BER performance
of the two schemes, showing that the BER for the macrocell users is
better with the scheme of Blind IA, whereas the BER performance in
the femtocells, which is 0 for the range of SNR values, is improved
with the hybrid scheme. Moreover, in the case of the hybrid scheme,
for high SNR values the BER performance of the two users in the macrocell
is almost the same, offering fairness and QoS to the edge-cell users.
Overall, looking at the total mean BER, the hybrid scheme outperforms
Blind IA.

\subsection{Rate Performance}

The rate of the network will be a function of the user's distance
from the base station and the amount of interference considered as
noise (in the case of the hybrid scheme). Again, users closer to the
transmitter can achieve higher rates compared to users at the edge
of the cell.

\begin{figure}
\begin{centering}
\begin{minipage}[t]{0.45\textwidth}%
\begin{center}
\includegraphics[width=6cm,height=5cm]{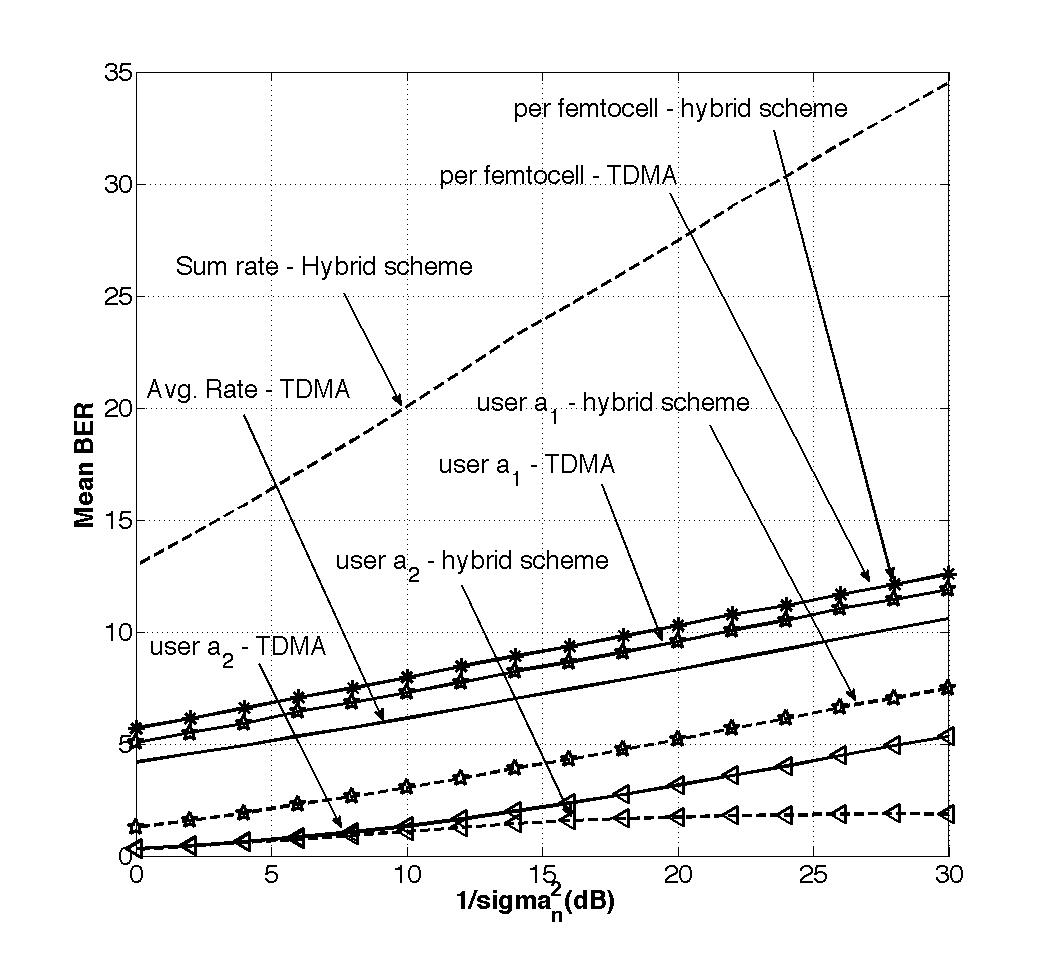}
\par\end{center}%
\end{minipage}\thinspace %
\begin{minipage}[t]{0.45\textwidth}%
\begin{center}
\includegraphics[width=6cm,height=5cm]{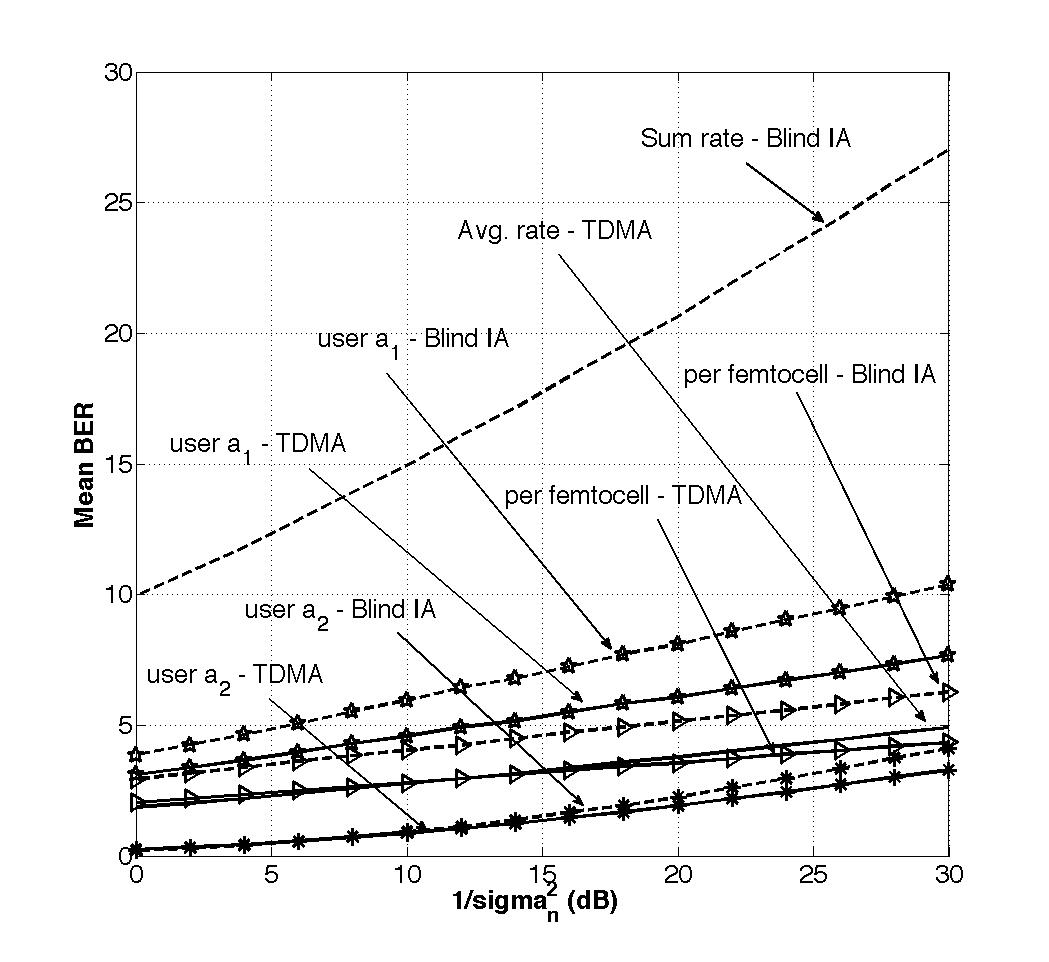}
\par\end{center}%
\end{minipage}
\par\end{centering}

\caption{Sum Rate Performance: (left) Hybrid vs. TDMA, (right) Blind IA vs.
TDMA. Both schemes achieve higher sum rates than TDMA \label{fig:Journal1-1-1-2-1-1-1-1-1} }
\end{figure}

Initially both schemes were compared to the case where only one user
is active (TDMA) using the following formulas for TDMA: \textbf{
\begin{equation}
R_{TDMA\,hybrid}^{[a_{k}]}=\frac{1}{T}\mathrm{\mathtt{\mathbb{E}}}\left[\log\det\left(\mathbf{I}_{N}+\frac{P_{\mathrm{macrocell}}}{N\sigma_{n}^{2}}\gamma_{a_{k}}\mathbf{h}_{a_{k}}\mathbf{h}_{a_{k}}^{T}\right)\right].\label{eq:RatemacroBIA-1}
\end{equation}
}

\begin{equation}
R_{TDMA\,hybrid}^{[f_{kl}]}=\frac{1}{T}\mathcal{\mathbb{E}}\left[\log\det\left(\mathbf{I}_{\mathcal{M}_{1}}+\frac{P_{\mathrm{femtocell}}}{N\sigma_{n}^{2}}\gamma_{f_{kl}}\mathbf{h}_{f_{kl}}\mathbf{h}_{f_{kl}}^{T}\right)\right].
\end{equation}
\textbf{
\begin{equation}
R_{TDMA\,BlindIA}^{[a_{k}]}=\frac{1}{T}\mathrm{\mathtt{\mathbb{E}}}\left[\log\det\left(\mathbf{I}_{N}+\frac{P_{\mathrm{macrocell}}}{N^{2}\sigma_{n}^{2}}\mathcal{D}^{[a_{k}]}\mathcal{K}^{[a_{k}]}\mathcal{K}^{[a_{k}]^{*}}\mathcal{D}^{[a_{k}]^{*}}\right)\right].\label{eq:RatemacroBIA-1-1}
\end{equation}
}

\begin{equation}
R_{TDMA\,BlindIA}^{[f_{kl}]}=\frac{1}{T}\mathcal{\mathbb{E}}\left[\log\det\left(\mathbf{I}_{\mathcal{M}_{1}}+\frac{P_{\mathrm{femtocell}}}{\mathcal{M}_{1}\sigma_{n}^{2}}\mathcal{K}^{[f_{kl}]}\mathcal{D}^{[f_{kl}]}\mathcal{D}^{[f_{kl}]^{*}}\mathcal{K}^{[f_{kl}]^{*}}\right)\right].
\end{equation}
Figure \ref{fig:Journal1-1-1-2-1-1-1-1-1} (left) depicts the BER
for every user separately, for both the hybrid and the TDMA schemes.
Note that the BER for users $a_{1}$, $f_{1}$ and $f_{2}$ is 0 for
the range of SNR values. Focusing on the total network BER for the
hybrid scheme and the average BER for the case of only one user in
the network being active, we can observe that both schemes offer similar
overall BER performances, with the average TDMA BER slightly outperforming
the hybrid scheme for high SNR values. Figure \ref{fig:Journal1-1-1-2-1-1-1-1-1}
(right) depicts the BER for every user separately, for both the hybrid
and the TDMA schemes. Focusing on the total network BER for the hybrid
scheme and the average BER for the case of only one user in network
being active, we can observe that both schemes offer similar overall
BER performances.

Figure \ref{fig:Journal1-1-1-2-1-1-1-1-2} (right) depicts the comparison,
in terms of every user's rate, between Blind IA and the hybrid scheme.
The rate of the macrocell users is better in the Blind IA case, however
femtocell users in the case of the hybrid scheme achieve higher rates. 

\section{Summary}

Overall, this paper introduces two novel management schemes for a
heterogeneous networks with $K$-users in the macrocell, and $KL$
femtocells, with $L$ femtocells interfering with every user in the
macrocell. The hybrid scheme provides power allocation fairness and
QoS to edge cell users, more DoF, and better performance to the femtocell
users, whereas Blind IA achieves considerably higher rates and lower
BERs for the users in the macrocell. Most importantly, both schemes
can achieve at least double the sum rate of TDMA, with the hybrid
scheme always achieving more DoF than TDMA. Due to the low system
overhead, high data rates, fair power allocation scheme and the heterogeneous
nature of both models, both schemes can be considered as candidates
for managing interference in 5G multi-tier communication networks,
depending on their requirements and architecture. Future work will
focus on wireless energy transfer and physical layer security, two
key aspects of future mobile networks, in network architectures that
employ the two proposed schemes.

\section{Acknowledgements}

This work was supported by NEC; the Engineering and Physical Sciences
Research Council {[}EP/I028153/1{]}; and the University of Bristol.

\section*{Appendix }
\begin{IEEEproof}
(Theorem 9) We show that $P^{[a_{k}]}$ removes intra- and inter-
cell interference at the \emph{$k$}th receiver. Substituting, (\ref{eq:MacrocellSignal})
and (\ref{eq:lloo}) in (\ref{eq:Paper1ast-1}), we consider coefficients
of $\mathbf{U}^{[a_{i}]}$ and $\mathbf{U}^{[f_{kl}]}$ separately.
For $i\neq k$, using $\left(A\otimes B\right)\left(C\otimes D\right)=\left(AC\right)\otimes\left(BD\right)$,
for intra-cell interference, coefficient of $\mathbf{U}^{[a_{i}]}$
becomes:
\begin{align}
\mathbf{P}^{[a_{k}]}\mathbf{H}^{[a_{k}]}\mathbf{V}^{[a_{i}]} & =\frac{a}{\sqrt{N}}{\displaystyle \sum_{s=1}^{2}}\left(\mathbf{w}_{s}^{[a_{k}]}\otimes\mathbf{D}_{s}^{[a_{k}]}\mathbf{\widetilde{h}}^{[f_{k}a_{k}]}\right)\sqrt{\gamma_{a_{k}}}(\mathbf{I}_{T}\otimes\mathbf{h}^{[a_{k}]})(\mathbf{v}^{[a_{i}]}\otimes\mathbf{I}_{N})\nonumber \\
 & =\frac{a}{\sqrt{N}}\sqrt{\gamma_{a_{k}}}{\displaystyle \sum_{s=1}^{2}}\left(\mathbf{w}_{s}^{[a_{k}]}\mathbf{v}^{[a_{i}]}\right)\otimes\left(\mathbf{D}_{s}^{[a_{k}]}\mathbf{\widetilde{h}}^{[f_{k}a_{k}]}\mathbf{h}^{[a_{k}]}\right),
\end{align}
where by 1) in Definition 8, for all s, $\mathbf{w}_{s}^{[a_{k}]}\mathbf{v}^{[a_{i}]}=0$,
i.e. $\mathbf{w}_{s}^{[a_{k}]}$ is orthogonal to $\mathbf{v}^{[a_{i}]}$
if $i\neq k$. For $i=k$ the remaining term is (\ref{eq:paper1-13-1}).
For inter-cell interference from Group $G_{1}$, coefficient of $\mathbf{U}^{[f_{kl}]}$:

\begin{align}
\mathbf{P}^{[a_{k}]}\mathbf{H}^{[f_{kl}a_{k}]}\mathbf{V}^{[f_{kl}]} & =\frac{b_{A}}{\sqrt{N}}{\displaystyle \sum_{s=1}^{2}}\left(\mathbf{w}_{s}^{[a_{k}]}\otimes\mathbf{D}_{s}^{[a_{k}]}\mathbf{\widetilde{h}}^{[f_{k}a_{k}]}\right)\sqrt{\gamma_{f_{kl}a_{k}}}\left(\mathbf{I}_{T}\otimes\mathbf{h}^{[f_{kl}a_{k}]}\right)\left(\sum_{i=1}^{T-1}\mathbf{\mathbf{\xi}}_{i}^{[f_{k}]^{T}}\otimes\mathbf{r}\mathbf{q}_{i}^{[f_{kl}]}\right)\nonumber \\
 & =\frac{b_{A}}{\sqrt{N}}\sqrt{\gamma_{f_{kl}a_{k}}}{\displaystyle \sum_{s=1}^{2}{\displaystyle \sum_{i=1}^{T-1}\mathbf{w}_{s}^{[a_{k}]}}}\mathbf{\xi}_{i}^{[f_{k}]^{T}}\otimes\mathbf{D}_{s}^{[a_{k}]}\mathbf{\widetilde{h}}^{[f_{k}a_{k}]}\mathbf{h}^{[f_{kl}a_{k}]}\mathbf{r}\mathbf{q}_{i}^{[f_{kl}]},
\end{align}
where for $\mathbf{s=1}$: the $(\mathbf{D}_{s}^{[a_{k}]}\mathbf{\widetilde{h}}^{[f_{k}a_{k}]}\mathbf{h}^{[f_{kl}a_{k}]}\mathbf{r})=0$.
Premultiplying by $\mathbf{D}_{s}^{[a_{k}]}$ selects a row of $\mathbf{\widetilde{h}}^{[f_{k}a_{k}]}$
and post multiplying by $\mathbf{r}$ selects a column of $\mathbf{\widetilde{h}}^{[f_{k}a_{k}]}$,
with the resulting row and column being orthogonal by 4) in Definition
8). For $\mathbf{s=2}$: the $(\mathbf{w}_{s}^{[a_{k}]}\mathbf{\xi}_{i}^{[f_{kl}]^{T}})=0$
by 2) in Definition 8. For inter-cell interference from Group $G_{2}$,
coefficient of $\mathbf{U}^{[f_{kl}]}$:
\begin{align}
\mathbf{P}^{[a_{k}]}\mathbf{H}^{[f_{kl}a_{k}]}\mathbf{V}^{[f_{kl}]} & =\frac{b_{B}}{\sqrt{N}}{\displaystyle \sum_{s=1}^{2}}\left(\mathbf{w}_{s}^{[a_{k}]}\otimes\mathbf{D}_{s}^{[a_{k}]}\mathbf{\widetilde{h}}^{[f_{k}a_{k}]}\right)\sqrt{\gamma_{f_{kl}a_{k}}}\left(\mathbf{I}_{T}\otimes\mathbf{h}^{[f_{kl}a_{k}]}\right)\left(\mathbf{v}^{[f_{kl}]T}\otimes\mathbf{e}_{2}\right)\nonumber \\
 & =\frac{b_{B}}{\sqrt{N}}\sqrt{\gamma_{f_{kl}a_{k}}}{\displaystyle \sum_{s=1}^{2}{\displaystyle \mathbf{w}_{s}^{[a_{k}]}}}\mathbf{v}^{[f_{kl}]T}\otimes\mathbf{D}_{s}^{[a_{k}]}\mathbf{\widetilde{h}}^{[f_{k}a_{k}]}\mathbf{h}^{[f_{kl}a_{k}]}\mathbf{e}_{2},
\end{align}
where for $\mathbf{s=1}$: the $(\mathbf{w}_{s}^{[a_{k}]}\mathbf{v}^{[f_{kl}]T})=0$
by 2) in Definition 8. For $\mathbf{s=2}$: the $(\mathbf{D}_{s}^{[a_{k}]}\mathbf{\widetilde{h}}^{[f_{k}a_{k}]}\mathbf{h}^{[f_{kl}a_{k}]}\mathbf{e}_{2})=0$.
Premultiplying by $\mathbf{D}_{s}^{[a_{k}]}$ selects a row of $\mathbf{\widetilde{h}}^{[f_{k}a_{k}]}$
and post multiplying by $\mathbf{e}_{2}$ selects a column of $\mathbf{\widetilde{h}}^{[f_{k}a_{k}]}$,
with the resulting row and column being orthogonal by 4) in Definition
8).
\end{IEEEproof}
\,
\begin{IEEEproof}
(Theorem 11) We show that $P^{[f_{kl}]}$ removes inter-cell interference
at the $kl$th receiver, so coefficient of $\mathbf{U}^{[a_{i}]}$
for all $i$, becomes:
\begin{align}
\mathbf{P}^{[f_{kl}]}\mathbf{H}^{[Af_{kl}]}\sum_{i=1}^{K}\mathbf{V}^{[a_{i}]} & =\frac{a}{\sqrt{N}}\left(\mathbf{w}\otimes\mathbf{W}^{[f_{kl}]}\right)\sqrt{\gamma_{Af_{kl}}}\left(\mathbf{I}_{T}\otimes\mathbf{h}^{[Af_{kl}]}\right)\left(\sum_{i=1}^{K}\mathbf{v}^{[a_{i}]}\otimes\mathbf{I}_{N}\right)\nonumber \\
 & =\frac{a}{\sqrt{N}}\sqrt{\gamma_{Af_{kl}}}\sum_{i=1}^{K}\left(\mathbf{w}\mathbf{v}^{[a_{i}]}\otimes\mathbf{W}^{[f_{kl}]}\mathbf{h}^{[Af_{kl}]}\right),
\end{align}
where by 1) in Definition 10, for all i, $\mathbf{w}\mathbf{v}^{[a_{i}]}=0$,
i.e. $\mathbf{w}$ is orthogonal to $\mathbf{v}^{[a_{i}]}$\emph{
}for all\emph{ }$i$. Coefficient of $\mathbf{U}^{[f_{kl}]}$ for
$l=2$, becomes:
\begin{align}
\mathbf{P}^{[f_{kl}]}\mathbf{H}^{[f_{k2}f_{kl}]}\mathbf{V}^{[f_{k2}]} & =\frac{b_{B}}{\sqrt{N}}\left(\mathbf{w}\otimes\mathbf{W}^{[f_{kl}]}\right)\sqrt{\gamma_{f_{k2}f_{kl}}}\left(\mathbf{I}_{T}\otimes\mathbf{h}^{[f_{k2}f_{kl}]}\right)\left(\mathbf{v}^{[f_{kl}]^{T}}\otimes\mathbf{e}_{2}\right)\nonumber \\
 & =\frac{b_{B}}{\sqrt{N}}\sqrt{\gamma_{f_{k2}f_{kl}}}\left(\mathbf{w}\mathbf{v}^{[f_{kl}]^{T}}\otimes\mathbf{W^{[f_{kl}]}}\mathbf{h}^{[f_{k2}f_{kl}]}\mathbf{e}_{2}\right),
\end{align}
where by 2) in Definition 10, $\mathbf{W}^{[f_{kl}]}\mathbf{h}^{[f_{k2}f_{kl}]}\mathbf{e}_{2}=0$.
\end{IEEEproof}
\,
\begin{IEEEproof}
(Theorem 13) We show that $P^{[f_{kl}]}$ removes inter-cell interference
at the $kl$th receiver, so coefficient of $\mathbf{U}^{[a_{i}]}$
for all $i$, becomes:
\begin{align}
\mathbf{P}^{[f_{kl}]}\mathbf{H}^{[Af_{kl}]}\sum_{i=1}^{K}\mathbf{V}^{[a_{i}]} & =\frac{a}{\sqrt{N}}\left(\mathbf{w}\otimes\mathbf{W}^{[f_{kl}]}\right)\sqrt{\gamma_{Af_{kl}}}\left(\mathbf{I}_{T}\otimes\mathbf{h}^{[Af_{kl}]}\right)\left(\sum_{i=1}^{K}\mathbf{v}^{[a_{i}]}\otimes\mathbf{I}_{N}\right)\nonumber \\
 & =\frac{a}{\sqrt{N}}\sqrt{\gamma_{Af_{kl}}}\sum_{i=1}^{K}\left(\mathbf{w}\mathbf{v}^{[a_{i}]}\otimes\mathbf{W}^{[f_{kl}]}\mathbf{h}^{[Af_{kl}]}\right),
\end{align}
where by 1) in Definition 12, for all i, $\mathbf{w}\mathbf{v}^{[a_{i}]}=0$,
i.e. $\mathbf{w}$ is orthogonal to $\mathbf{v}^{[a_{i}]}$\emph{
}for all\emph{ }$i$. Coefficient of $\mathbf{U}^{[f_{kl}]}$ for
$l=1,...,L$ and $l\neq2$, becomes.
\begin{align}
\mathbf{P}^{[f_{kl}]}\mathbf{H}^{[f_{kl}f_{k2}]}\mathbf{V}^{[f_{kl}]} & =\frac{b_{B}}{\sqrt{N}}\left(\mathbf{w}\otimes\mathbf{W}^{[f_{kl}]}\right)\sqrt{\gamma_{f_{kl}f_{k2}}}\left(\mathbf{I}_{T}\otimes\mathbf{h}^{[f_{kl}f_{k2}]}\right)\left(\sum_{i=1}^{T-1}\mathbf{\mathbf{\xi}}_{i}^{[f_{k}]^{T}}\otimes\mathbf{r}\mathbf{q}_{i}^{[f_{kl}]}\right)\nonumber \\
 & =\frac{b_{B}}{\sqrt{N}}\sqrt{\gamma_{f_{kl}f_{k2}}}\left(\sum_{i=1}^{T-1}\mathbf{w}\mathbf{\xi}^{[f_{kl}]^{T}}\otimes\mathbf{W^{[f_{kl}]}}\mathbf{h}^{[f_{kl}f_{k2}]}\mathbf{r}\mathbf{q}_{i}^{[f_{kl}]}\right),
\end{align}
where by 2) in Definition 12, $\mathbf{W}^{[f_{kl}]}\mathbf{h}^{[f_{kl}f_{k2}]}\mathbf{r}=0$
for $l=1,...,L$ and $l\neq2$.\end{IEEEproof}

\end{document}